\documentclass[twocolumn,nofootinbib,noshowpacs,preprintnumbers,amsmath,amssymb,secnumarabic]{revtex4}
\usepackage{graphicx, epsfig, natbib}

\DeclareRobustCommand{\SkipTocEntry}[4]{}

\newcommand{\be}{\begin{equation}}
\newcommand{\ee}{\end{equation}}
\newcommand{\bea}{\begin{eqnarray}}
\newcommand{\eea}{\end{eqnarray}}

\def\beq{\begin{equation}}
\def\eeq{\end{equation}}
\def\be{\begin{equation}}
\def\ee{\end{equation}}
\def\bea{\begin{eqnarray}}
\def\eea{\end{eqnarray}}
\def\d{{\rm d}}

\begin{document}
\title{Phenomenology of D-Brane Inflation with General Speed of Sound}

\author{Hiranya Peiris,$^1$\footnote{Hubble Fellow}
Daniel Baumann,$^2$ Brett Friedman,$^3$ and Asantha Cooray$^3$}

\affiliation{
$^1$Kavli Institute for Cosmological Physics and 
Enrico Fermi Institute, University of Chicago, Chicago, IL 60637\\
$^2$Department of Physics, Princeton University,Princeton, NJ 08544 \\
$^3$Center for Cosmology, Department of Physics and Astronomy, 
University of California, Irvine, CA 92697-4575}

\begin{abstract}
A characteristic of D-brane inflation is that
fluctuations in the inflaton field can propagate at a speed significantly
less than the speed of light.
This yields observable effects that are distinct from those of 
single-field slow roll inflation, such as a modification of the inflationary consistency relation and a potentially large level of non-Gaussianities.
We present a numerical algorithm that extends the inflationary flow formalism to models with general speed of sound.
For an ensemble of D-brane inflation models parameterized by the Hubble parameter and the speed of sound as polynomial functions of the inflaton field, we give qualitative predictions for the key inflationary
observables.
We discuss various consistency relations for D-brane inflation, 
and compare the qualitative shapes of the warp factors we derive from the numerical models with 
analytical warp factors considered in the literature. Finally, we derive and apply
a generalized microphysical bound on the inflaton field variation during brane inflation. While a large number of 
models are consistent with current
cosmological constraints, almost all of these models violate the compactification
constraint on the field range in four-dimensional Planck units. 
If the field range bound is to hold, then models with a detectable level of non-Gaussianity predict a blue scalar spectral index, and a tensor component that is far below the detection limit 
of any future experiment.
\end{abstract}


\maketitle
\tableofcontents
\vfil
\section{Introduction}

Understanding the physics of inflation \cite{Inflation} is one of the main challenges of fundamental physics and modern cosmology. Since string theory remains the most promising candidate for a UV completion of the Standard Model that unifies gauge and gravitational interactions in a consistent quantum theory, it seems natural to search within string theory for an explicit realization of the inflationary scenario. 
This search has so far revealed two distinct classes of inflationary models which identify the inflaton field with either open string modes ({\it e.g.} brane inflation \cite{Dvali, KKLMMT, otherbraneinflation}, DBI inflation \cite{DBI}), or closed string modes ({\it e.g.} K\"ahler moduli inflation \cite{Kahler}, racetrack inflation \cite{racetrack}, N-flation \cite{Nflation}). 
The brane inflation scenario \cite{Dvali, KKLMMT} in particular has received considerable theoretical and phenomenological interest. At the same time, precise cosmological observations \cite{Observations} have made the detailed predictions of inflation testable.\\

In this paper we study the general phenomenology of D-brane inflation models with arbitrary speed of sound. The theoretical ansatz is motivated by brane--anti-brane dynamics in warped spaces as described by the Dirac-Born-Infeld (DBI) action. Relativistic dynamics of the brane motion 
leads to 
deviations of the propagation speed of inflaton fluctuations from the speed of light. These effects generically result in a large level of non-Gaussianity of primordial fluctuations \cite{Chen} and modify the inflationary single-field consistency relation \cite{Lidsey_Seery}. We study these exciting observational signatures using a generalization of the inflationary flow formalism. This allows us to explore an ensemble of D-brane inflation models parameterized by the evolution of the Hubble parameter and the speed of sound as  polynomial functions of the inflaton field.\\

The outline of the paper is as follows: In \S\ref{sec:background} we review the DBI mechanism of D-brane inflation and discuss its basic cosmological predictions.
In \S\ref{sec:micro} we discuss microscopic consistency constraints that these models have to satisfy. 
In particular, we review and generalize the field range bound of \cite{BauMcA}. 
In \S\ref{sec:flow} we apply the inflationary flow formalism to D-brane inflation. We derive the generalized flow equations and use them to study the phenomenology of an ensemble of brane inflation models with arbitrary speed of sound. \S\ref{sec:results} contains a summary of our main results. We conclude in \S\ref{sec:conclusion}.
An Appendix gives further details of the Monte-Carlo algorithm.
We use natural units throughout the paper, where $c=\hbar \equiv 1$ and $M_P^{-2} \equiv 8 \pi G$.
Interesting related work has appeared in \cite{Bean} and \cite{Lidsey}.

\begin{table*}[!ht]
\begin{center}
\begin{tabular}{cll}
\hline
Variable &  Description & Notes\\
\hline \hline
$\gamma(\phi)$ & Inverse sound speed &  equations~(\ref{equ:gamma}) and (\ref{equ:g}) / Monte-Carlo \\
$H(\phi)$ & Hubble parameter & Monte-Carlo\\
$\epsilon$, $\eta$, $\kappa$ & Slow variation parameters & equations~(\ref{equ:eps}) to (\ref{eq:alphahier})  \\
$P_s(k)$ & Scalar power spectrum & We use $k_{\rm CMB}=0.02\, {\rm Mpc}^{-1}$\\
  $P_t(k)$ & Tensor power spectrum & $P_t = r P_s$ \\
$r$ & Tensor-to-scalar ratio & equation~(\ref{equ:r}); derived at $k_{\rm CMB}$\\
$n_s$ & Scalar spectral index & equation~(\ref{equ:ns}); derived at $k_{\rm CMB}$\\
$n_t$ & Tensor spectral index & equation~(\ref{equ:nt}); derived at $k_{\rm CMB}$\\
$f_{NL}$ & NG parameter & equation~(\ref{equ:fNL}); DBI prediction.\\
$N_e$ & Number of $e$-folds & $\d N_e \equiv - H \d t$\\
\hline
$\chi$ & Radial coordinate & equation~(\ref{equ:throat}) \\
$\varphi$ & Canonical inflaton & $\varphi^2 = T_3 r^2$ \\
$\phi$ & Monte Carlo inflaton & $\phi =0$ at the UV end of the throat. \\
$T_3$ & D3-brane tension & equation (\ref{equ:T3}) \\
$h^{-1}(\varphi)$ & Warp factor & equation (\ref{equ:lineelement}) and \S\ref{sec:micro}\\
$f^{-1}(\varphi)$ & Warped brane tension & $f^{-1} \equiv T_3 h^{-1}$. Same as $T(\varphi)$ in Ref.~\cite{Bean}\\
$g_s$ & String coupling & $g_s < 1$ \\
$\alpha'$ & String scale & $\alpha' M_P^2 \propto V_6^w$; equation~(\ref{equ:Ms}) \\
$M_P$ & 4d Planck mass & $M_P = m_{Pl}/\sqrt{8 \pi}$ \\
$M$, $K$ & Flux on $A$- and $B$-cycles & integer quantum numbers \\
$N$ & Five-form flux & $N \equiv MK \gg 1$\\
\hline
\end{tabular}
\end{center}
NOTES.---%
A summary of variables and functions related to our description of DBI inflation and observables.
\end{table*}

\section{DBI Inflation}
\label{sec:background}

\subsection{Review of Warped D-brane Inflation}

Since a quantum theory of (super)strings is naturally defined in ten spacetime dimensions, any realistic description of cosmology and particle phenomenology requires compactification of six of the space dimensions. Different compactification geometries lead to different low energy effective theories in four dimensions. In addition, the extra dimensions have to be stabilized, since even a small time-dependence in the size and shape of the extra dimensions can manifest itself in a variation of fundamental coupling constants and induce observable fifth forces.

Recently, Kachru, Kallosh, Linde and Trivedi (KKLT) \cite{KKLT} provided a framework for stabilizing the extra dimensions of type IIB string theory in the presence of background fluxes and non-perturbative effects (see \cite{FluxReview} for a review of flux compactification). The flux background fixes the {\it shape} (complex structure moduli) of the extra dimensions, but leaves the overall {\it size} (K\"ahler moduli) unfixed \cite{GKP}.  As \cite{KKLT} showed, the size of the compact space may be stabilized by the inclusion of non-perturbative effects, {\it e.g.}~gaugino condensation on D7-branes. 
In addition to stabilizing the shape of the extra dimensions, the background fluxes lead to strong warping of the spacetime. This warping provides an elegant mechanism to produce exponential hierarchies (a realization of the Randall-Sundrum scenario in string theory \cite{RS}) and 
has important consequences for the dynamics of brane motion in the warped space.

The line element of a warped flux compactification of type IIB string theory
to four dimensions takes the following form 
\beq
\label{equ:lineelement} \d s^2 = h^{-1/2}(y) g_{\mu\nu} \d x^{\mu} \d
x^{\nu} + h^{1/2}(y)g_{ij} \d y^{i} \d y^{j} \, , \eeq 
where $h$ is the warp factor and $\mu, \nu = 0, \dots , 3$ and $i,j = 4, \dots , 9$. 
Typically, the internal space will have one or more conical
throats sourced by the background fluxes, {\it{i.e.}} regions in which the metric is locally of the
form
\beq
\label{equ:throat}
g_{ij}\d y^{i} \d y^{j} = \d \chi^2 + \chi^2 \d s_{X_5}^2\, ,\eeq
for some five-manifold $X_5$ which forms the base of the cone. 
If the background contains suitable fluxes, the metric in the throat
region can be highly warped, $h^{-1} \ll 1$.\\

The KKLMMT scenario \cite{KKLMMT} refers to brane inflation \cite{Dvali} in a warped throat of a KKLT flux compactification \cite{KKLT}. 
In particular, most models of brane inflation are concerned with the dynamics of a mobile D3-brane that fills four-dimensional spacetime and is point-like in the compact space.
Here, the inflaton field $\varphi$ is identified with the geometrical separation between a D3-brane and an anti-D3-brane. The anti-brane is fixed at the tip of the throat ($\chi \equiv \chi_{\rm IR} \approx 0$), while the brane moves from large radius ($\chi \le \chi_{\rm UV}$) to small radius ($\chi_{\rm IR}$).\footnote{In this paper we consider the so-called 'UV-model', {\it cf.} \cite{ChenIR} for an interesting alternative where the brane moves out of the throat.}
The dynamics of a D3-brane in the warped background (\ref{equ:lineelement}) is governed by the Dirac-Born-Infeld (DBI) action
\beq \label{equ:actionP}
S = \frac{1}{2} \int \d^4 x \sqrt{-g} \Bigl[ M_P^2 {\cal R} + 2
{\cal P}(X, \varphi)\Bigr]\, , \eeq where  
\beq
\label{equ:dbiaction} {\cal P}(X,\varphi) \equiv - f^{-1}(\varphi) \sqrt{1-
2 f(\varphi) X} + f^{-1}(\varphi) - V(\varphi)\, .\eeq
Here $\varphi = \sqrt{T_3} \chi$ is the inflaton field, $T_3$ is the tension of the D3-brane and
$f^{-1}(\varphi) = T_3 h^{-1}(\varphi)$ is the rescaled warp factor of the background spacetime. 
 For a homogeneous background $X \equiv - \frac{1}{2}
g^{\mu \nu} \partial_\mu \varphi \partial_\nu \varphi \approx \frac{1}{2} \dot{\varphi}^2$ is the canonical kinetic term of the inflaton.
The potential for the brane motion $V(\varphi)$ arises from moduli stabilization effects \cite{KKLMMT, BDKMMM} and the Coulombic brane--anti-brane interaction.
In the slow-roll limit $f X \ll 1$, the DBI action (\ref{equ:dbiaction}) reduces to the following familiar form
 \beq {\cal P}(X, \varphi) \equiv X - V(\varphi)\, . \eeq
This slow-roll limit can be understood as the non-relativistic motion of the brane in the presence of a weak force from a flat potential.
The relativistic limit of brane motion in a warped background is characterized by the parameter $\gamma$ (defined in analogy to the Lorentz factor of relativistic particle dynamics)
\beq
\label{equ:gamma}
\gamma \equiv \frac{1}{\sqrt{1 - f(\varphi) \dot{\varphi}^2}}\, .
\eeq
Positivity of the argument of the square-root in (\ref{equ:gamma}) imposes a local speed limit on the brane motion, $\dot \varphi^2 \le f^{-1}(\varphi) = T_3 h^{-1}(\varphi)$. The presence of strong warping, $h^{-1} \ll 1$, in the throat can make this maximal speed of the brane much smaller than the speed of light. When $\dot \varphi$ is close to this speed limit, then $\gamma$ is large.

From the inflaton action~(\ref{equ:actionP})
we find the homogeneous energy density in the field 
\bea
\rho &=& 2 X
{\cal P}_{,X}-{\cal P} \nonumber \\
&=& (\gamma -1) f^{-1} + V\, ,
\eea
while the pressure is
\beq
{\cal P} = (1- \gamma^{-1}) f^{-1} - V\, .
\eeq
$\rho$ and ${\cal P}$ source the dynamics of the homogeneous background spacetime, $\d s^2 = -\d t^2 + a(t)^2 \d {\mathbf x}^2$, as described by the Friedmann equations
\bea
3 M_P^2 H^2 &=& \rho\\
2 M_P^2 \dot{H} &=& -(\rho +  {\cal P})\, .
\eea
Accelerated expansion ($\ddot a > 0$) requires smallness of the variation of the Hubble parameter $H \equiv \partial_t \ln a$, as quantified by the parameter
\beq
\epsilon \equiv - \frac{\dot{H}}{H^2} = \frac{3}{2} (1 + w) < 1\, ,
\eeq
where
\beq
\label{equ:w}
w \equiv \frac{{\cal P}}{\rho} = \frac{(1- \gamma^{-1}) f^{-1} - V} {(\gamma -1) f^{-1} + V}< - \frac{1}{3}\, .
\eeq
From the expression for the equation of state parameter (\ref{equ:w}) we see that although the brane moves relativistically in the DBI limit, $\gamma \gg 1$, inflation still requires that the potential energy $V$ dominates over the kinetic energy of the brane $(\gamma -1) T_3 h^{-1}
$. This is possible because the kinetic energy of the brane is suppressed by the large warping of the internal space, $h^{-1} \ll 1$.

As discussed in Ref. \cite{Tye}, the slow roll (non-relativistic brane motion) and the DBI (ultra-relativistic brane motion) limits are connected continuously by an intermediate regime where the relativistic effects are small, $\gamma = {\cal O}(1)$, but non-negligible.

\subsection{Cosmological Observables}
\label{sec:cosmo}

To discuss the phenomenological predictions of the original DBI inflationary scenario \cite{DBI} and its generalizations \cite{Tye}, it is convenient to define the speed of sound 
\beq c_s^2 \equiv
\frac{d{\cal P}}{d\rho} = \frac{{\cal P}_{,X}}{\rho_{,X}} =
\frac{{\cal P}_{,X}}{{\cal P}_{,X}+2X {\cal P}_{,XX}} = \frac{1}{\gamma^2}\, . \eeq
This is the speed at which fluctuations of the inflaton $\delta \varphi$ propagate relative to the homogeneous background.
In addition, we define slow variation
parameters in analogy with the standard slow roll parameters for inflation with canonical kinetic term \bea
\label{equ:epsilon}
\epsilon &\equiv& - \frac{\dot H}{H^2} = \frac{X {\cal P}_{,X}}{M_P^2 H^2}\, ,\\
\tilde \eta &\equiv & \frac{\dot \epsilon}{\epsilon H}\, , \label{equ:tildeeta}\\
\kappa &\equiv& \frac{\dot{c_s}}{c_s H}\, . \eea 

\addtocontents{toc}{\SkipTocEntry}
\subsubsection{\sl Perturbation Spectra}

To first order in
the slow variation parameters, the basic cosmological observables as a function of wavenumber $k$
are \cite{Garriga} \bea
P_s(k) &=& \left. \frac{1}{8 \pi^2 M_P^2} \frac{H^2}{c_s \epsilon} \right|_{k c_s = aH}\, ,\\
P_t(k) &=& \left. \frac{2}{\pi^2} \frac{H^2}{M_P^2} \right|_{k = aH}\, .
\eea
Scalar perturbations freeze when they exit the sound horizon $k c_s = a H$, while tensor perturbations freeze when $k = a H$.
Over a limited range of scales it is appropriate to parameterize deviations from perfect scale-invariance by the following spectral indices
\bea
n_s -1 &\equiv& \frac{d P_s}{d \ln k} =  - 2 \epsilon - \tilde \eta - \kappa  \label{equ:ns}\, ,\\
n_t &\equiv& \frac{d P_t}{d \ln k} = -2 \epsilon\, . \label{equ:nt} \eea 
Notice the dependence of $n_s$ on the evolution of the speed of sound as captured by the parameter $\kappa$.
The tensor-to-scalar ratio in these models is 
\beq \label{equ:r} r \equiv \frac{P_t}{P_s} = 16\,  c_s
\epsilon\, . \eeq The dependence of (\ref{equ:r}) on the speed of
sound implies a modified consistency relation 
\beq 
\label{equ:mce}
r = - 8 \, c_s n_t\, . 
\eeq 
As discussed recently by Lidsey and
Seery \cite{Lidsey_Seery}, equation
(\ref{equ:mce}) provides an interesting possibility of testing
DBI inflation. 

The standard slow roll predictions for the cosmological perturbation spectra
are recovered in the limit $c_s \to 1$, $\dot c_s \to 0$.

\addtocontents{toc}{\SkipTocEntry}
\subsubsection{\sl Non-Gaussianity}

The non-trivial structure of the kinetic term in the action (\ref{equ:dbiaction}) leads to striking observational signatures of relativistic DBI inflation. In particular, it was observed \cite{DBI} that this generically leads to a very large non-Gaussianity of the primordial fluctuations.

Let us give a brief qualitative description of the physical origin of this non-Gaussianity\footnote{The following discussion parallels the treatment in Ref. \cite{ClineReview}}
 before citing the results of an exact computation \cite{Chen}.
Consider the unperturbed kinetic term
\beq
{\cal L}_{\rm kin} = - f^{-1} \sqrt{1-f \dot \varphi^2} =
- \frac{f^{-1}}{\gamma}
\eeq 
and its first order variation under $\varphi \to \varphi + \delta \varphi$
\beq
\delta^{(1)} {\cal L}_{\rm kin} =  \gamma \dot \varphi \, \delta \dot \varphi\, . \eeq
This indicates that self-couplings of $\varphi$ are enhanced by factors of $\gamma$ arising from expansion of the square root of the DBI action.
Non-Gaussianities come from the third-order interactions in ${\cal L}_{\rm kin}$ due to the perturbation $\delta \varphi(t, \mathbf{x})$.
The leading effect can be estimated by considering the ratio of the cubic perturbation to the matter Lagrangian, $\delta^{(3)} {\cal L}_{\rm kin}$, to the quadratic perturbation, $\delta^{(2)} {\cal L}_{\rm kin}$, neglecting mixing with gravitational perturbations. Since matter self-interactions must dominate in order to obtain significant non-Gaussian features this provides a rough estimate of the effect.
In the large $\gamma$ limit the leading contribution to the non-Gaussianity parameter $f_{NL}$ scales as \cite{DBI}
\beq
f_{NL} \propto \frac{\delta^{(3)}{\cal L}_{\rm kin}}{\delta^{(2)} {\cal L}_{\rm kin}} \propto \gamma^2\, .
\eeq
We observe that the magnitude of non-Gaussianities scales with the Lorentz factor $\gamma$. Observational constraints on primordial non-Gaussianities therefore lead to interesting constraints on the magnitude of relativistic DBI effects.

The non-Gaussianity of fluctuations in DBI inflation was estimated in the large $\gamma$ limit in \cite{DBI} and computed more precisely for the general case in \cite{Chen}. 
Observational tests of the non-Gaussianity of the primordial
density perturbations are most sensitive to the three-point
function of the comoving curvature perturbations $\zeta$. It is usually
assumed that the three-point function has a form that would follow
from the field redefinition \beq \zeta = \zeta_g - \frac{3}{5}
f_{NL} \zeta_g^2\, , \eeq where $\zeta_g$ is Gaussian. The scalar
parameter $f_{NL}$ then quantifies the amount of non-Gaussianity. It is a function of three momenta which form a triangle in Fourier space. Here we cite results for the limit of an equilateral triangle. The general shape of non-Gaussianities in DBI inflation may be found in \cite{Chen}.  Slow
roll models predict \cite{Juan} $f_{NL} \ll 1$, which is far
below the detection limit of present and future observations. 
For generalized inflation models represented by the action (\ref{equ:actionP}) with general pressure function ${\cal P}$, one finds \cite{Chen}
\beq
\label{equ:fNL}
f_{NL} = \frac{35}{108} \Bigl( \frac{1}{c_s^2} -1 \Bigr) - \frac{5}{81} \Bigl( \frac{1}{c_s^2} -1 - 2 \Lambda\Bigr)\, ,
\eeq
where
\beq
\Lambda \equiv \frac{X^2 {\cal P}_{,XX} + \frac{2}{3} X^3 {\cal P}_{,XXX}}{X {\cal P}_{,X} + 2 X^2 {\cal P}_{,XX}}\, .
\eeq
For the specific case of DBI inflation (\ref{equ:dbiaction}) the second term in (\ref{equ:fNL}) is identically zero \cite{Chen} and the
 prediction for the level of non-Gaussianity  is
\beq
\label{equ:fNL2}
f_{NL} = \frac{35}{108} \Bigl( \frac{1}{c_s^2} - 1\Bigr) \approx
\frac{1}{3} (\gamma^2 -1)\, .\eeq 
Measurements of primordial non-Gaussianity therefore constrain the speed of sound during the time when CMB scales exit the horizon during inflation.
This implies an upper bound on $\gamma$ from
the observed upper limit on the non-Gaussianity of the primordial
perturbations.  Using the {\it WMAP} limit \cite{fNL}, $-256 < f_{NL}
< 332$ (95\% confidence level),
one finds
$\gamma < 32$.
Ref.~\cite{MZ06} forecasts constraints on primordial non-Gaussianity from {\it Planck} of $f_{NL}^{\rm equil} <66.9$ at the $1\sigma$ level, which translates to $\gamma < 14$.

Using (\ref{equ:fNL2}) the modified consistency relation (\ref{equ:mce}) can be expressed completely in terms of observables \cite{Lidsey_Seery}
\beq
8 n_t = - r \sqrt{1 + \frac{108}{35} f_{NL}}\, .
\eeq
Although not inconceivable, it will be challenging to experimentally test this unique prediction of brane inflation.
For the standard slow-roll case with $8n_t=-r$, the combination of next-generation CMB polarization measurements of $r$ and estimates of $n_t$ based on the direct detection of gravitational waves fails to test this consistency relation to better than 50\% for even the most optimistic experimental scenarios \cite{Smith:06}.

\section{Microscopic Constraints}
\label{sec:micro}

The philosophy of this paper is to take maximal theoretical guidance from the string theoretic origin of D-brane inflation models, but use a phenomenological Ansatz to capture the largest set of possible scenarios in a single framework. In particular, we allow considerable freedom on the functional form of the inflaton potential $V(\varphi)$ and the background warp factor $h(\varphi)$. Ultimately, both $V$ and $h$ should of course be derived from an explicit string compactification (see \cite{BDKM}). Here, we take the approach of studying a general set of functional forms for these quantities, restricted only by cosmological constraints (\S\ref{sec:cosmo}) and a minimal set of microscopic consistency requirements. 
In this section, we describe the microscopic constraints that we impose on our models.

\subsection{Warped Background}

\addtocontents{toc}{\SkipTocEntry}
\subsubsection{\sl Warped Geometry} 

Schematically, the warp factor in equation~(\ref{equ:lineelement}) is determined by the solution of the Laplace equation
\beq
\label{equ:h}
-\nabla^2 h \sim |G_3|^2\, ,
\eeq
where $\nabla^2$ is the six-dimensional Laplacian and $G_3$ parameterizes 3-form flux which sources the warping.
Intuitively, equation (\ref{equ:h}) is just like the equation for an electrostatic potential on the compact Calabi-Yau space in the presence of a `charge density'.
The functional form of the warp factor $h$ hence depends on how the `charge density' $|G_3|^2$ is distributed.
If it is localized in one region, then $h$ will tend to decrease
monotonically away from it. We call this the one-throat situation.
$h$ should reach some finite value ({\it i.e.} it should not diverge) at the bottom of the throat, since otherwise the warped string and brane tensions which scale as $h^{-1}$ would vanish. 
In addition, to avoid a naked singularity $h$ should not vanish.

The Klebanov-Strassler (KS) geometry \cite{KS} is an explicit non-compact ten-dimensional solution to type IIB supergravity in the presence of background fluxes. The KS spacetime decomposes into the form (\ref{equ:lineelement}) with   
the internal space (\ref{equ:throat}) given by  a cone over $X_5 = T^{1,1}$.
Far from the tip of the throat the 
KS solution is well approximated by $AdS_5 \times T^{1,1}$ with the warp factor determined by the Green's function of (\ref{equ:h})
\beq
\label{equ:AdS}
h_{AdS} = \left(\frac{R}{\chi} \right)^4\, , \qquad \chi > \chi_{\rm IR}\, ,
\eeq
where
\beq
\frac{R^4}{(\alpha')^2} =4 \pi g_s N \frac{\pi^3}{{\rm Vol}(X_5)}\, , \quad N \equiv M K\, .
\eeq
Here, ${\rm Vol}(X_5)$ parameterizes the dimensionless volume of $X_5$ with unit radius.
Typically, ${\rm Vol}(X_5) = {\cal O}(\pi^3)$, {\it e.g.} ${\rm Vol}(T^{1,1}) = \frac{16 \pi^3}{27}$.
$g_s < 1$ and $l_s \equiv \sqrt{\alpha'}$ are the string coupling and the string length, respectively and $\chi_{\rm IR}$ is the minimal radial coordinate at the tip of the throat,
\beq
\ln (R/\chi_{\rm IR}) \approx \frac{2 \pi K}{3 g_s M}\, .
\eeq
The integers $K$ and $M$ denote flux quanta on the $A$ and $B$ cycles of the throat.
The exact KS warp factor is non-singular at the tip \cite{KS}.
The AdS warp factor (\ref{equ:AdS}) forms the basis for many theoretical studies of brane motion in warped throat regions. However, other warped solutions are possible (and some are known), so in the spirit of our phenomenological approach and to retain maximal generality we allow $h(\chi)$ to be a free function subject only to minimal theoretical constraints.  
Of course, we appreciate that it is therefore {\it not} guaranteed that all (or even most) warp factors we consider in this study have explicit microscopic realizations.
More detailed theoretical constraints on the functional form of $h(\chi)$ than the ones presented in this section are beyond the scope of this paper. For a more complete theoretical understanding of the phenomenology of DBI inflation, such a theoretical study is essential.

If one considers a scenario with two (or more) throats, corresponding to two localized `charge' distributions in (\ref{equ:h}), then $h$ will not be monotonic overall (similarly
to the electric potential of two positive charges). It will reach a minimum
between the throats (`charges') where $h \approx 1$.
However, within each throat it should behave monotonically, as
in the explicitly known examples of gauge/gravity duality like the KS solution. Let us remark that 
although we will focus our attention to single throat scenarios with monotonic warp factors,
the methodology of the present paper is easily generalized to studies of multi-throat scenarios.

\addtocontents{toc}{\SkipTocEntry}
\subsubsection{\sl Warped Brane Tension}

The dynamics of a D3-brane in a warped throat region is determined by the brane potential $V(\varphi)$ and the warped brane tension $f^{-1}(\varphi) = T_3 h^{-1}(\varphi)$. In the following we relate
the scale of the warped tension
to microscopic parameters.
The D3-brane tension in four-dimensional Planck units is
\beq
\label{equ:T3}
\frac{T_3}{M_P^4} = \frac{1}{(2 \pi)^3} \frac{1}{g_s} \left( \frac{M_s}{M_P} \right)^4 \, ,
\eeq
where $M_s^{-2} \equiv \alpha'$ defines the string scale.
The string scale is related to the four-dimensional Planck mass (or the Newton constant $G$) via the (warped) compactification volume $V_6^w\equiv \int \d^6 y \sqrt{g}\, h$
\beq
\label{equ:Ms}
\left( \frac{M_P}{M_s}\right)^2 = \frac{2}{(2 \pi)^7} \frac{1}{g_s^2} \frac{V_6^w}{(\alpha')^3}\, .
\eeq
Notice that the four-dimensional Planck mass increases as the (warped) volume of the internal manifold is increased. 
The warp factor in the throat region $h^{-1}$ lies in the following range $[h^{-1}_{\rm IR}, h^{-1}_{\rm UV}]$, where
\beq
h_{\rm IR}^{-1} \approx e^{- 8 \pi K/ 3g_s M} 
\eeq
is the warp factor at the tip of the throat and $h_{\rm UV}^{-1} \approx 1$ defines the region where the throat is glued into a bulk space. Small string coupling and large $K/M$ allow exponentially large warping, {\it i.e.} small $h^{-1}_{\rm IR}$.
Using $f^{-1} = T_3 h^{-1}$, we find
\bea
f M_P^4 &=& (2 \pi)^3 g_s \left( \frac{M_P}{M_s} \right)^4 h\\
&=& \frac{4}{(2 \pi)^{11}} \frac{1}{g_s^3} \left( \frac{V_6^w}{l_s^6} \right)^2 h \, .
\eea
To facilitate comparison with the analysis of Bean {\it et al.} \cite{Bean}, 
let us consider the same fiducial values for the string coupling and the string scale,  $g_s = 0.1$, $(M_P/M_s)^2 = 1000$.
This implies fixing the D3-brane tension to have the following value, $T_3 \approx 10^{-8} M_P^4$.
Allowing the warp factor $h^{-1}$ to be in the range $h^{-1} \in [10^{-10},1]$ then implies $f M_P^4 \in [10^8,10^{18}]$ for the fiducial range of $f$.
However, we note that by decreasing the string scale (increasing the volume of the internal space) and/or increasing the warping, one can make $f$ much larger.
In general, we treat the magnitude of $f$ as a free parameter of the models.

\subsection{Bound on the Field Range}
 
Since the inflaton field $\varphi$
for D-brane inflation acquires a geometrical meaning, there exist geometrical restrictions on the allowed range of $\varphi$. In particular, the finite size of the compact extra dimensions restricts the field variation of the canonical inflaton in four-dimensional Planck units. Naively, since $\varphi \propto \chi$, one might imagine that the inflaton field range can be increased by simply scaling up the radial dimension of the throat. However, this increases the volume of the compact space, and by equation (\ref{equ:Ms}), changes the four-dimensional Planck mass. For the warped cones that form the basis of most explicit theoretical models, the inflaton field range in four-dimensional Planck units in fact decreases as the radial dimension of the cone is increased.

More specifically, recall from (\ref{equ:T3}) and (\ref{equ:Ms}) that the four-dimensional Planck mass scales with the {\it warped} volume of the compact space
\beq
M_P^2 = \frac{1}{\pi} (T_3)^2 V_6^w\, . 
\eeq
Using the conservative bound that the total warped volume (which receives contributions from the throat and the bulk) is at least as big as the warped volume of the throat only, {\it i.e.}
\beq
V_6^w > (V_6^w)_{\rm throat} = {\rm Vol}(X_5) \int_{\chi_{\rm IR}}^{\chi_{\rm UV}} \d \chi \, \chi^5 h(\chi)\, ,
\eeq
one finds
\beq
\label{equ:PlanckMass}
M_P^2 > \frac{{\rm Vol}(X_5)}{\pi} \int_{\varphi_{\rm IR}}^{\varphi_{\rm UV}} \d \varphi \, \varphi^5 f(\varphi)\, .
\eeq
For the cut-off AdS warp factor
\beq
\label{equ:AdS2}
f_{AdS} = \frac{\lambda}{\varphi^4} \, , \quad \quad \lambda \equiv T_3 R^4 = \frac{\pi}{2} \frac{N}{{\rm Vol}(X_5)}\, ,
\eeq
this implies that
\beq
M_P^2 > \frac{N}{4} \varphi_{\rm UV}^2\, ,
\eeq
and hence
\beq
\left(\frac{\varphi_{\rm UV}}{M_P}\right)^2 < \frac{4}{N}\, .
\eeq
Since $\Delta \varphi \le \varphi_{\rm UV}$, these microscopic considerations imply the following bound on the total field variation during warped D-brane inflation \cite{BauMcA} 
\beq
\label{equ:dphi}
\left( \frac{\Delta \varphi}{M_P} \right)^2 < \frac{4}{N}\, ,
\eeq
where for theoretical consistency the flux integer $N=M K$ has to be much greater than unity.
In all explicitly known examples, super-Planckian field variations are therefore microscopically disallowed and brane-inflation models that predict $\Delta \varphi > M_P$ should be viewed with suspicion. 

For ultra-relativistic DBI inflation ($f_{NL} > 1$) with a quadratic potential one can show that the normalization of the primordial scalar spectrum requires \cite{BauMcA}
\beq
\label{equ:lower}
N = \left( \frac{32}{3 \pi} \right)^3 \frac{3\, {\rm Vol}(X_5)}{(r^2 f_{NL})^2 P_s} > 10^8 {\rm Vol}(X_5).
\eeq
Since typically ${\rm Vol}(X_5) = {\cal O}(\pi^3)$, observations therefore imply very large $N$, allowing only very small field variations. 
However, Ref.~\cite{BauMcA} also derived an upper limit on $N$ in terms of the observational limits on non-Gaussianity and tensors
\beq
\label{equ:upper}
N < \frac{27}{70} r f_{NL} < 40\, .
\eeq
The limits (\ref{equ:lower}) and (\ref{equ:upper}) are clearly inconsistent unless ${\rm Vol}(X_5)$ is extremely small.

The numerical analysis of Bean {\it et al.}~\cite{Bean} shows that the intuition gained from the explicit example of \cite{BauMcA} extends to the non-analytic, intermediate DBI regime. In particular, Bean {\it et al.} find that imposing the microscopic bound (\ref{equ:dphi}) dramatically reduces the parameter space of viable DBI models in the intermediate and ultra-relativistic regime. 
In this paper we study if this conclusion continues to hold when $f(\varphi)$ is allowed to be a free function.

For general $f(\varphi)$, the total field variation during inflation is bounded by $\Delta \varphi < \varphi_{\rm UV}$, or
\beq
\label{equ:dphi2}
\left( \frac{\Delta \varphi}{M_P}\right)^2 < \frac{4}{N_{\rm throat}}\, ,
\eeq
where
\beq
N_{\rm throat} \equiv 
\frac{ 4\,  {\rm Vol}(X_5) }{\pi \, \varphi_{\rm UV}^2}  \int_{\varphi_{\rm IR}}^{\varphi_{\rm UV}} \d \varphi \, \varphi^5 f(\varphi)\, .
\eeq
Typically, $N_{\rm throat} \gg 1$.
In this paper we compute $N_{\rm throat}$ by direct numerical integration of our output warp factors $f(\varphi)$.
Models that violate (\ref{equ:dphi2}) cannot be embedded in a consistent string compactification.
We find that most models discussed in this paper violate this condition, which highlights the theoretical challenge of constructing microscopically viable models of UV DBI inflation that are consistent with observations.

\subsection{Microscopic Bound on Tensors}

The field range bound has important implications for the expected level of gravitational waves from brane inflation models \cite{BauMcA}.
By the Lyth bound \cite{LythBound}, the constraint on the evolution of the inflaton (\ref{equ:dphi}), or the generalized bound (\ref{equ:dphi2}), is related to the maximal observable gravitational wave signal from inflation. Let us recall the argument and generalize it to general speed of sound theories.
Restricting to the homogeneous mode $\varphi(t)$ 
we find from (\ref{equ:epsilon})  
that \beq \frac{\d \varphi}{M_P} = - \sqrt{\frac{2
\epsilon}{{\cal P}_{,X}} } \, \d {N_e} \eeq and hence \beq \label{equ:int}
\frac{\Delta \varphi}{M_P}= \int_{N_{\rm end}}^{{N}_{\rm CMB}}
\sqrt{\frac{r}{8} \frac{\gamma}{{\cal P}_{,X}}} \, \d {N_e}\, , \eeq
where $\Delta \varphi \equiv \varphi_{\rm CMB} - \varphi_{\rm end}$.
Notice the non-trivial generalization of the standard slow roll result
through the factor ${\cal P}_{,X}/\gamma$. For DBI
inflation this factor happens to be unity, since ${\cal P}_{,X} = \gamma$,
so that the Lyth bound
remains the same as for slow roll inflation
\beq
\label{equ:lyth}
r_{\rm CMB} = \frac{8}{N_{\rm eff}^2} \left( \frac{\Delta \varphi}{M_P} \right)^2\, ,
\eeq
where
\beq
N_{\rm eff} \equiv \int_{N_{\rm end}}^{N_{\rm CMB}} \d N_e \left(\frac{r}{r_{\rm CMB}} \right)^{1/2}
\eeq
quantifies the support of the integral (\ref{equ:int}) and $r_{\rm CMB}$ denotes the tensor-to-scalar ratio evaluated on CMB scales.
The value of $N_{\rm eff}$ depends on the evolution of $r(N_e)$ after CMB scales have exited the horizon during inflation. 
In slow roll models this evolution is typically small (second order in slow roll), with observational data on CMB scales requiring $N_{\rm eff} = {\cal O}(50-60)$, leading to the prediction of an unobservably small level of gravitational waves, $r_{\rm CMB} \ll 0.01$ \cite{BauMcA}.

\subsection{Implications for Relativistic DBI}

Lidsey and Huston \cite{Lidsey} derived an interesting generalization of the field range bound of \cite{BauMcA} that is useful in the DBI limit ($f_{NL} \gg 1$). Here we briefly review their argument.
Let $\Delta \varphi_\star$ be the field variation when observable scales are generated during inflation, corresponding to $\Delta N_\star \le 4$ $e$-foldings of inflationary expansion (this corresponds to the CMB multipole range $2 \le \ell < 100$). 
Then the integral in the bound on the Planck mass (\ref{equ:PlanckMass}) can be approximated as follows
\beq
\int_{\varphi_{\rm IR}}^{\varphi_{\rm UV}} \d \varphi \varphi^5 f(\varphi) > \Delta \varphi_\star \varphi_\star^5 f_\star > (\Delta \varphi_\star)^6 f_\star\, .
\eeq
Here, we have bounded the integral by a small part of the Riemann sum, defined $f_\star \equiv f(\varphi_\star)$ and used $\Delta \varphi_\star < \varphi_\star$.
Equation (\ref{equ:PlanckMass}) then becomes
\beq
\left( \frac{\Delta \varphi_\star}{M_P}\right)^6 < \frac{\pi}{{\rm Vol}(X_5)} (f_\star M_P^4)^{-1}\, .
\eeq
Next, we note that the warped tension $f^{-1}$ can be expressed in terms of 
the scalar power spectrum $P_s$, the tensor-to-scalar ratio $r$, and the non-Gaussianity parameter $f_{NL}$. 
$f_\star$ can therefore be related to observables
\beq
(f_\star M_P^4)^{-1} = \frac{\pi^2}{16} P_s^\star r_\star^2 \left( 1 + \frac{1}{3 f_{NL}} \right)\, ,
\eeq
and hence
\beq
\label{equ:lidsey}
\left( \frac{\Delta \varphi_\star}{M_P}\right)^6 < \frac{\pi^3}{ 16 {\rm Vol}(X_5)}  P_s^\star r_\star^2 \left( 1 + \frac{1}{3 f_{NL}} \right)\, .
\eeq
For slow roll models with $f_{NL} \ll 1$ this is not a very useful constraint.
However, for relativistic DBI models with $f_{NL} \gg 1$ the bound (\ref{equ:lidsey}) is independent of $f_{NL}$. Since, $P_s^\star \sim 2.4 \times 10^{-9}$, $r_\star < 0.5$ (from observations) and ${\rm Vol}(X_5) = {\cal O}(\pi^3)$ (from theory) we conclude that super-Planckian field variation is inconsistent with observations.
The Lyth bound (\ref{equ:lyth}) can now be written as
\beq
\left(\frac{\Delta \varphi_\star}{M_P}\right)^2 \approx {r_\star \over 8 }(\Delta N_\star)^2\,.
\eeq
Substituting this into (\ref{equ:lidsey}) we find \cite{Lidsey}
\beq
\label{equ:rlidsey}
r_\star < \frac{32 P_s^\star}{(\Delta N_\star)^6} \frac{\pi^3}{{\rm Vol}(X_5)}\, .
\eeq
The observed level of scalar fluctuations $P_s^\star \sim 2.4 \times 10^{-9}$ therefore implies that the tensor amplitude is unobservably small for relativistic DBI models if the field range bound of \cite{BauMcA} is applied (unless ${\rm Vol}(X_5)$ is made unnaturally small). We emphasize that the bound (\ref{equ:rlidsey}) does {\it not} apply to slow roll models since $f_{NL} > 1$ has been assumed in its derivation.

Interestingly, Lidsey and Huston also derived a lower limit on $r$ for models of UV DBI inflation with $f_{NL} \gg 1$ and $n_s < 1$ \cite{Lidsey}
\beq
\label{equ:rlidsey2}
r_\star >  \frac{4 (1-n_s)}{\sqrt{3 f_{NL}}}\, .
\eeq
The limits (\ref{equ:rlidsey}) and (\ref{equ:rlidsey2}) are clearly inconsistent unless ${\rm Vol}(X_5)$ is very small, {\it cf.}~(\ref{equ:lower}) and (\ref{equ:upper}).\\

With the work of \cite{BauMcA}, \cite{Bean} and \cite{Lidsey} there is now a growing body of evidence that the best motivated theoretical models of relativistic (UV) DBI inflation are in tension with the data if microscopic constraints are applied consistently.
In this paper we reach conclusions that are consistent with this and show that these problems persist even if considerable freedom is allowed for the functional form of the brane potential $V(\varphi)$ and the background warp factor $f(\varphi)$.

\section{The Flow Formalism for D-brane Inflation}
\label{sec:flow}

To study the ensemble of inflationary models specified by the theoretical ansatz of the previous section, we adapt the inflationary flow formalism \cite{Hoffman_Turner, Kinney, Easther_Kinney}. In this section we derive the generalized flow equations for brane inflation. In the next section we present our numerical results.

\subsection{Inflation in the Hamilton-Jacobi Approach}

Let us recall the Hamilton-Jacobi (HJ) approach to inflationary dynamics and apply it to warped brane inflation. 
For a spatially flat FRW spacetime,
the scale factor $a(t)$ is determined by the Friedmann equation
\beq
\label{equ:Friedmann}
H^2 = \frac{1}{3 M_P^2} \rho\, ,
\eeq
where $\rho = (\gamma -1)f^{-1} + V$ and $\gamma(\phi) \equiv [1-f(\phi) \dot{\phi}^2]^{-1/2}$ follow from the DBI action (\ref{equ:dbiaction}).
 The inflaton\footnote{In the following we use the variable $\phi$ for the inflaton field rather than $\varphi$ as in section 2, since we want to allow for the possibility that $\phi=0$ in our Monte Carlo simulation doesn't coincide with the tip of the throat.
 In fact, $\phi=0$ will be the UV end of the throat. The variables $\varphi$ and $\phi$ are simply related by a linear transformation corresponding to this shift in origin. } 
 $\phi$ obeys the following equation of motion 
\beq
\label{equ:eom}
\ddot{\phi} + \frac{3 f'}{2f} \dot{\phi}^2 - \frac{f'}{f^2} + \frac{3 H}{\gamma^2} \dot{\phi} + \Bigl( V' + \frac{f'}{f^2}\Bigr) \frac{1}{\gamma^3} = 0\, ,
\eeq
where primes denotes derivatives with respect to $\phi$.
In the HJ-formalism the Hubble expansion rate $H(\phi)$ is considered the fundamental quantity.
The master equation relating $\phi(t)$ and $H(\phi)$ follows from equations (\ref{equ:Friedmann}) and (\ref{equ:eom})
\beq
\label{equ:1}
H'(\phi)\equiv \frac{d H}{d \phi} = \frac{\dot H}{\dot \phi} = - \gamma(\phi)\, \frac{\dot \phi}{2 M_P^2}\, .
\eeq
From equation (\ref{equ:1}) one finds
\beq
\label{equ:g}
\gamma(\phi) = \sqrt{1 + 4 f(\phi) M_P^4 \left[H^\prime(\phi)\right]^2}\, .
\eeq
or
\beq
\label{equ:HJ0}
f(\phi)^{-1}  = \frac{4 [H']^2 }{\gamma^2 -1} M_P^{4}\, .
\eeq
Given $H(\phi)$ and $\gamma(\phi)$, the inflaton potential is
\beq
\label{equ:HJ}
V(\phi) = 3 M_P^2 H^2 - 4 M_P^4 \frac{[H']^2}{\gamma + 1}\, .
\eeq
Notice the following important consequence of the Hamilton-Jacobi equations (\ref{equ:HJ0}) and (\ref{equ:HJ}): For any specified function $H(\phi)$ and $\gamma(\phi)$, it produces a potential $V(\phi)$ and a warp factor $f(\phi)$ which admits the given $H(\phi)$ and $\gamma(\phi)$ as an exact inflationary solution.

\subsection{Inflationary Flow Equations}

Analogous to the Hubble slow roll (HSR) parameters, we can write down a set of slow variation
parameters for brane inflation that are defined in terms of derivatives of $H$ and $\gamma$ with respect to $\phi$:
\begin{eqnarray}
\epsilon(\phi) &\equiv& \frac{
 2 M^2_P}{\gamma(\phi)} \left( \frac{H'(\phi)}{H(\phi)}  \right)^2 \label{equ:eps} \, ,\\
\eta(\phi) &\equiv& \frac{ 2 M^2_P}{\gamma(\phi)}\frac{H''(\phi)}{H(\phi)}  \, ,\\
\kappa(\phi) &\equiv& \frac{2 M^2_P}{\gamma(\phi)}\frac{H'(\phi)\gamma'(\phi)}{H(\phi)\gamma(\phi)}  \, ,\\
^{\ell}\lambda(\phi) &\equiv& \left(\frac{2 M^2_P}{\gamma(\phi)}\right)^\ell  \left( \frac{H'}{H} \right)^{\ell-1}\frac{1}{H(\phi)} \frac{d^{\ell+1}H(\phi)}{d\phi^{\ell+1}} \label{eq:lambdahier}\, , \\ 
^{\ell}\alpha(\phi) &\equiv& \left(\frac{2 M^2_P}{\gamma(\phi) }\right)^{\ell}  \left( \frac{H'}{H} \right)^{\ell-1}\frac{1}{\gamma(\phi)} \frac{d^{\ell+1}\gamma(\phi)}{d\phi^{\ell+1}} \label{eq:alphahier}  \, , 
\end{eqnarray}
where $\ell \geq 1$. For notational convenience we define $\eta \equiv {}^{1}\lambda$, $\xi \equiv {}^2\lambda$, $\rho \equiv {}^{1}\alpha $ and
$\sigma \equiv {}^{2}\alpha$.
Note that $\eta$ is related to $\tilde \eta$ in (\ref{equ:tildeeta}) by $\tilde \eta =2 \epsilon - 2 \eta - \kappa$.
The trajectories of these parameters are governed by a set of
coupled first order differential equations. 
Using the relation 
\begin{equation}
\label{equ:phiN}
\frac{d \phi}{d N_e} = + \frac{2 M_P^2}{\gamma} \frac{H'}{H} \, , \quad \d N_e \equiv - H \d t\, ,
\end{equation}
we find
\begin{eqnarray}
\label{equ:s1}
\frac{d\epsilon}{dN_e}&=&-\epsilon (2\epsilon-2 \eta+\kappa)\, ,  \\
\frac{d\eta}{dN_e}&=&-\eta(\epsilon+\kappa)+\xi \, , \\
\frac{d^{\ell}\lambda}{dN_e}&=&-^{\ell}\lambda\,  \Bigl[\ell \kappa+ \ell \epsilon-(\ell-1)\eta \Bigr] +
\ ^{\ell+1}\lambda\, , \label{equ:s3} 
\end{eqnarray}
and
\begin{eqnarray}
\label{equ:ss1}
\frac{d\kappa}{dN_e}&=&-\kappa\, (2\kappa+\epsilon-\eta)+\epsilon\rho\, , \\
\frac{d\rho}{dN_e}&=&-2\rho \kappa+\sigma \, , \\
\frac{d^{\ell}\alpha}{dN_e}&=&-^{\ell}\alpha\, \Bigl[(\ell+1)\kappa+(\ell-1)\epsilon \nonumber \\
&&-(\ell-1)\eta \Bigr] +\  ^{\ell+1}\alpha\, . \label{equ:ss3}
\end{eqnarray}
This set of differential equations defines the flow equations for brane inflation. 
Notice that while the flow parameters depend on $\gamma$,
the flow equations do not depend on $\gamma$ explicitly.

\begin{figure*}[]
\psfig{file=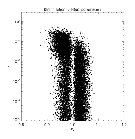,width=3.5in}\hfill
\psfig{file=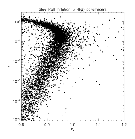,width=3.5in}
\caption{The $r-n_s$ plane populated by numerical models of DBI inflation with $M=5$ and $M'=1$ in equations~(\ref{eq:h}) and (\ref{eq:gamma}), combining simulations with general and small $\gamma_0$ priors (left panel) and by numerical models of standard slow-roll inflation from Kinney \cite{Kinney:2002qn} (right panel).}
\label{fig:KinneyPlot}
\end{figure*}

As pointed out by Liddle \cite{Liddle_flow}, the flow equations have an analytic solution. 
Truncating the hierarchy of flow equations so that the last non-zero terms are $^{M}\lambda$
and $^{M'}\alpha$ ensures that $^{M+1}\lambda$=$0$ and $^{M'+1}\alpha$=$0$ at all times 
(along with all the higher order terms). 
From Eqs.~(\ref{eq:lambdahier}) and (\ref{eq:alphahier}), it then follows that higher derivatives vanish at all times:
\begin{eqnarray}
\frac{d^{(M+2)} H}{d\phi^{(M+2)}} &=& 0 \, ,\\
\frac{d^{(M'+2)} \gamma}{d\phi^{(M'+2)}} &=& 0\, .
\end{eqnarray}
We thus arrive at polynomials of order $M+1$ and $M'+1$ in $\phi$ respectively for the functions $H(\phi)$ and $\gamma(\phi)$,
\begin{eqnarray}
H(\phi) &=&  H_0\Biggl[ 1+ A_1 \left(\frac{\phi}{M_P}\right) +  \dots  \nonumber \\
	&& \hspace{1cm}  \dots +  A_{M+1} \left(\frac{\phi}{M_P}\right)^{M+1}\Biggr], \label{eq:h}  \\
\gamma(\phi) &=& \gamma_0 \Biggl[ 1+ B_1 \left(\frac{\phi}{M_P}\right) + \dots \nonumber \\
	&& \hspace{1cm} \dots + B_{M'+1} \left(\frac{\phi}{M_P}\right)^{M'+1} \Biggr].  \label{eq:gamma} 
\end{eqnarray}
So far, we haven't specified an initial value for $\phi$.   If we truncate the series as above, any set of flow parameters spans an $M+1$ dimensional space. However, the evolution equations define the flow parameters as functions of $\phi$. Consequently, the set of distinct  trajectories spans an $M$ dimensional space, effectively fibering the space of initial conditions for the flow hierarchy. However, if the flow parameters are specified at $\phi=0$ the ambiguity is removed. 

From the definitions of the flow parameters, the coefficients $A_j$ and $B_j$ can be written 
in terms of the initial values of the flow parameters 
\begin{eqnarray}
A_1 &=& \sqrt{\epsilon_0 \gamma_0 / 2}\, , \\
A_{\ell+1} &=& \frac{\left(\gamma_0/2\right)^\ell }{(\ell+1)! 
\ A_1^{\ell-1}} \  {}^{\ell}\lambda_{0}\, , \label{eq:Acoeffs}
\end{eqnarray}
and
\begin{eqnarray}
B_1 &=& \frac{\kappa_0 \gamma_0 / 2}{A_1}\, , \\
B_{\ell+1} &=& \frac{ (\gamma_0/2)^{\ell} }{(\ell+1)! 
\ A_1^{\ell-1}} \ {}^{\ell}\alpha_{0} \, . \label{eq:Bcoeffs} 
\end{eqnarray}
The sign convention we choose is as follows. To define the direction of time with respect to the number of $e$-folds before the end of inflation $N_e$, we choose $N_e$ to increase as one goes {\em backward} in time; {\it i.e.}, $\d N_e = -H \d t$, so that $\d t > 0$ as $\d N_e <0$. 
Furthermore, we choose $ \sqrt{\epsilon}$ to have the same sign as $H'(\phi)$. This is equivalent to choosing $A_1>0$ in our notation; the sign of $A_1$ specifies in which direction the field is rolling. \\

Note that in our present work $f(\phi)$ is not fixed to be a specific function. Instead, $f(\phi)$ is a derived quantity that is determined through equation~(\ref{equ:HJ0}) via Monte-Carlo descriptions of $\gamma(\phi)$ and $H(\phi)$ . Similarly, $V(\phi)$ is also a derived function determined through $H(\phi)$ and $\gamma(\phi)$ using equation~(\ref{equ:HJ}). In this sense, our work differs significantly from the recent work of Bean {\it et al.} \cite{Bean} in that they fix both $V(\phi)$ and $f(\phi)$ to particular (well-motivated) forms.
We allow a large range of warp-factors through a specification of a general sound speed captured by $\gamma(\phi)$, and a large range of dynamics captured by $H(\phi)$. Thus, the observable distributions shown in Bean {\it et al.} \cite{Bean} for quantities such as the scalar spectral index and the tensor-to-scalar ratio are under a fixed theoretical model whose internal parameters (such as the normalization of the AdS warp factor) are allowed to vary. This can be considered a `top-down' approach. Our work adopts the complementary `bottom-up' viewpoint that, while some (or even many) of our warp factors may not have microscopic realizations, it is still interesting  to  investigate the phenomenology of the set of warp factors, potentials, and observables that are generally allowed by existing cosmological data.

\begin{figure*}[]
\psfig{file=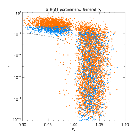,width=3.5in}\hfill
\psfig{file=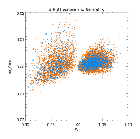,width=3.5in}\\
\psfig{file=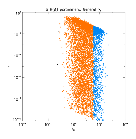,width=3.5in}\hfil
\psfig{file=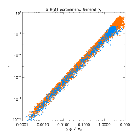,width=3.5in}
\caption{Cosmological observables evaluated at $k_{\rm CMB}$ for a simulation applying the general $\gamma_0$ prior with $M=5$ and $M'=1$ in equations~(\ref{eq:h}) and (\ref{eq:gamma}). The three regimes of DBI inflation are shown with color-coded points: slow roll DBI (black), intermediate DBI (orange) and ultra-relativistic DBI (blue). The individual plots from top-left in clockwise direction are: (a) the scalar-to-tensor ratio $r$ vs. scalar spectral index $n_s$, (b) spectral index $n_s$ vs. running of the spectral index $\alpha_s=dn_s/d\ln k$, (c) $r$ vs. $\Delta \phi= \phi_0-\phi_{\rm end}$, and (d) $r$ vs. non-Gaussianity parameter $f_{\rm NL}$. In (a), we show the lower limit on $r$ for ultra-relativistic DBI models with $r > (1-n_s)/8$ from \cite{Lidsey} as a dashed line. In (d), the limit from WMAP, $|f_{\rm NL}| <332$ \cite{fNL} is indicated by a vertical dashed line.
In (c), the relation between $r$ and $\Delta \phi$ can be described approximately as $r \propto (\Delta \phi/M_P)^2$. This follows from equation (\ref{equ:lyth}); integration from $N_0$ to $N_{\rm end}$ instead of from $N_{\rm CMB}$ to $N_{\rm end}$ just results in a different normalization $N_{\rm eff}$.}
\label{fig:2}
\end{figure*}

\begin{figure*}[]
\psfig{file=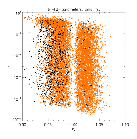,width=3.5in}\hfill
\psfig{file=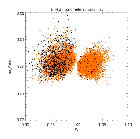,width=3.5in}\\
\psfig{file=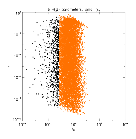,width=3.5in}\hfill
\psfig{file=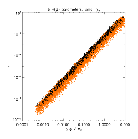,width=3.5in}
\caption{Same as Fig. \ref{fig:2}, but with a small $\gamma_0$ prior. Note that while in Fig.  \ref{fig:2}(a), the tensor/scalar ratio for ultra-relativistic DBI models satisfied a general lower bound, this is not the case for non-relativistic models.}
\label{fig:3}
\end{figure*}

\subsection{Monte-Carlo Algorithm for DBI Inflation} \label{subsec:algorithm}

Now we shall outline the algorithm used to produce the numerical results of \S\ref{sec:results}; technical details of the specific implementation are described in Appendix A.

The algorithm is designed to capture the dynamics of DBI inflation from the time the brane enters the mouth of the throat and inflation begins, until inflation ends (either through a tachyonic transition near the tip of the throat or due to slow roll ending). Because we explicitly define the direction of motion as  the brane moving into the throat ($\phi$ decreases), the algorithm as described below only applies to {\it monotonic} warp factors (single throat scenarios). However, it is straightforward to generalize the method to include the brane moving out of the throat ($\phi$ increases) and thereby model multi-throat scenarios with non-monotonic warp factors; we plan to explore this possibility in future work.

A key feature of the algorithm is a well-defined physical ``location'' for $\phi=0$, corresponding to the mouth of the throat, $\phi_{\rm UV}$. Here and in the following, the subscript $``0"$ denotes evaluation at $\phi = 0$.

There are three phenomenological classes of DBI inflation that we would like to consider in this work. We classify models by their value of $\gamma$ at CMB scales: 
\begin{enumerate}
\item {\sf slow roll DBI}: $1< \gamma_{\rm CMB} < 1.1$ 
\item {\sf intermediate DBI}: $1.1< \gamma_{\rm CMB} < 10$
\item {\sf ultra-relativistic DBI}: $\gamma_{\rm CMB} > 10$. 
\end{enumerate}
We are particularly interested in the limit of the DBI classes above which are consistent with the microphysical bound (\ref{equ:dphi2}) on the field range. In order to sample these different regimes efficiently, we consider different combinations of priors on $\gamma_0$ and $\epsilon_0$ in our Monte-Carlo simulations, as follows:
\begin{itemize}
\item {\sf  Small $\gamma_0$ initial conitions}: We assume that the initial speed of the brane is very small, so
\beq
\gamma_0 = \gamma(0) \equiv 1+\Delta\, ,
\eeq
for $\Delta \ll 1$
and draw $\log_{10} \gamma_0$ randomly from the narrow flat prior $ [0,10^{-5}]$. This prior is effective at sampling the slow roll and intermediate DBI regimes well. 
\item  {\sf General $\gamma_0$ initial conditions}: We relax the above assumption on the initial speed of the brane, drawing $\log_{10} \gamma_0$ randomly from the broad flat prior $ [0,1]$. This prior is effective at sampling the intermediate and ultra-relativistic DBI regimes well. 
\item {\sf General $\epsilon_0$ initial conditions}: We allow a key observable quantity, the tensor-to-scalar ratio, to take a wide range of values by drawing from the broad flat prior $\log_{10} \epsilon_0 \in [-10,0]$. Unless specified otherwise, this is the standard prior we will use in \S \ref{sec:results}.
\item {\sf Small $\epsilon_0$ initial conditions}: As we will find below, the general $\epsilon_0$ prior generates DBI models that are consistent with the current cosmological constraints, but which strongly violate the microphysical bound (\ref{equ:dphi2}). In order to quantify the limit on the tensor-to-scalar ratio at which the field range bound is satisfied for the different classes of models, we draw from the narrow flat prior $\log_{10} \epsilon_0 \in [-21,-9]$.
\end{itemize}

The rest of the algorithm is identical for all the cases that we consider. We Monte-Carlo over $P_0 \equiv P_s (\phi=0)$, the initial amplitude of the power spectrum of scalar density perturbations at $\phi = 0$ and its value at the CMB scale $P_s(k_{\rm CMB})$. The initial values at the UV end for a limited number of the slow variation parameters in equations~(\ref{eq:h}, \ref{eq:gamma}), which one expects may be constrained by current and future cosmological data, are also picked randomly. The priors for these initial conditions are specified in \S A.1. The initial conditions at $\phi=0$ fix $f_0$ and $H_0$,
\bea
f_0 &=& \frac{\gamma_0^2-1}{16 \pi^2 \epsilon_0^2 P_0} M_P^{-4}\, , \\ 
H_0 &=& \sqrt{\frac{\gamma_0^2-1}{2 \epsilon_0 \gamma_0 f_0 M_P^4}} M_P \, .
\eea
We numerically evolve the flow equations forward into the throat ($\phi$ decreases) until we find a match to the scalar power spectrum amplitude at CMB scales: 
\beq
\frac{\gamma_{\rm CMB}}{8 \pi^2 \epsilon_{\rm CMB}} \left( \frac{H_{\rm CMB}}{M_P} \right)^2 = P_s(k_{\rm CMB}) \label{eq:cmbmatch1}
\eeq
The way that the matching condition at CMB scales is implemented is described in detail in \S A.2. Once this condition is satisfied, we can compute the observable quantities by linking the value of the inflaton field $\phi$ to the comoving wavenumber $k(\phi)$ of cosmological perturbations, as follows.

Without loss of generality, we pick some fiducial physical scale $k_{\rm CMB}$ to correspond to $\phi=\phi_{\rm CMB}$. Then, with our sign convention, $\phi>\phi_{\rm CMB}$ corresponds to scales larger than $k_{\rm CMB}$, and $\phi<\phi_{\rm CMB}$ corresponds to smaller scales. 
The wavenumber of cosmological perturbation modes is associated with a value of $\phi$ through \cite{Tye, Garriga}
\begin{equation}
\frac{dN_e}{d\ln k} = -\frac{1}{1-\epsilon-\kappa} \label{eq:phiNeq} \, ,
\end{equation}
so
\begin{equation}
\frac{d\frac{\phi}{M_P}}{d\ln k} = - \sqrt{\frac{2 \epsilon}{\gamma}} \frac{1}{1-\epsilon-\kappa} \label{eq:phikeq} \, ,
\end{equation}
where the last expression follows from equation (\ref{equ:phiN}). Hence we associate a wavenumber with a value of $\phi$ by solving Eq.~(\ref{eq:phikeq}).

Instead of the amplitude of the power-spectra at each $k$, the observables are widely described in terms of the power-law variables. To second order in slow variation parameters the spectral indices of the scalar and tensor perturbations are \cite{Tye, Bean}
\begin{eqnarray}
n_s - 1 &\equiv& \frac{d P_s}{d \ln k} = (1 + \epsilon + \kappa) (-4 \epsilon + 2 \eta - 2 \kappa)\, , \\
n_t &\equiv& \frac{d P_t}{d \ln k} = \frac{- 2 \epsilon}{1 - \epsilon - \kappa}\, .
\end{eqnarray}
We also consider the variation with scale, or ``running'', of the spectral indices
\begin{eqnarray}
\alpha_s &\equiv& \frac{dn_s}{d \ln k} = \frac{4{d\epsilon \over dN_e}-2 {d\eta \over dN_e}+2 {d\kappa \over dN_e} + \cdots}{(1-\epsilon-\kappa)^2}\, ,\\
\alpha_t &\equiv& \frac{dn_t}{d\ln k}=\frac{2 {d\epsilon \over dN_e} + \cdots}{(1-\epsilon-\kappa)^2}\, .
\end{eqnarray}

In our simulations we compute the cosmological observables $P_s(k)$, $P_t(k)$, and $f_{NL}$, as well as the various power-law variables, at
$k_{\rm CMB}=0.02$  Mpc$^{-1}$  \cite{Cortes:2007ak, Peiris:2006sj}  using the $k(\phi)$ equation~(\ref{eq:phikeq}). Cosmological scalar modes freeze as they exit the {\it sound horizon}, $c_s H^{-1} =(\gamma H)^{-1}$, while tensor modes freeze when their scale exceeds the Hubble radius, $H^{-1}$. 
Therefore, $k_{\rm CMB}$ corresponds to different values of the field  $\phi_s$ and $\phi_t$ for the scalar and tensor modes respectively, and this difference is taken into account when computing observables. For the tensor/scalar ratio, we calculate $P_t(\phi_t)/P_s(\phi_s)$, rather than using the analytic expression for $r$. 

In addition, throughout the evolution, we impose a set of constraints to enforce consistency of the physical picture, which we describe in detail in \S A.2. 

Now we will consider the implementation of the end of inflation in our numerical models, leaving a detailed description until \S A.3. In most models brane inflation ends via a tachyonic instability as the separation between a D3-brane and an anti-D3-brane becomes comparable to a string length. This correlates with the warp factor reaching a minimal value since the anti-brane minimizes its energy at the tip of the throat where the warp factor is smallest. Here, we consider two distinct scenarios:
\begin{enumerate}
\item Inflation ends by tachyonic instability. 
\item Inflation ends before the tachyonic instability sets in, {\it i.e.}~$\epsilon \to 1$ before the D3-brane comes within one string length of the anti-brane. The tachyonic mode then only serves to remove the inflationary energy density, so that the vacuum has zero energy after reheating.
\end{enumerate}

One more ingredient is needed to complete the description of our algorithm. We found that simply Monte-Carlo'ing the priors for initial slow-roll parameters as outlined in \S A.1 produced large numbers of models which, while satisfying all our physical constraints, had observables which were very different from the current cosmological data. This is because the evolution of flow-equations is very sensitive to the initial conditions due to the rapid variations in the function $\gamma(\phi)$, which when combined with $H(\phi)$ leads to rapid variations in $f(\phi)$. While the goal of this present work is not to make a detailed comparison of the model with the data, we are nevertheless primarily interested in the properties of generalized DBI models whose evolution histories are broadly consistent with current observations.

In order to preferentially increase the population of such models in our simulations, we apply the following selection mechanism: a  Metropolis-Hastings algorithm (which is used in standard Markov Chain Monte Carlo techniques) is implemented with a ``penalty function" taken from the scalar $P(k)$ results of Fig.~10 (bottom right panel) from Ref.~\cite{Peiris:2006sj}. This figure contains a scalar $P(k)$ reconstructed from WMAP 3 year data \cite{Hinshaw:2006ia, Page:2006hz} and the SDSS galaxy power spectrum \cite{Tegmark:2003uf}, under the assumption that the primordial fluctuations are seeded by the standard single field slow roll inflation mechanism that additionally satisfies a minimal ``sufficient $e$-folds'' requirement that solves the cosmological flatness and horizon problems \cite{Peiris:2006ug}. Since we expect, in general, that the present model, which contains more parameters, will be less well constrained by the current data, the Metropolis-Hastings algorithm is run at a high ``temperature''. In practice, this means that the penalty function, a least-square statistic, uses \emph{double} the 95\% CL error of the figure as a 1--$\sigma$ Gaussian error, allowing the DBI model considerably more freedom to deviate from the mean than the single field slow roll case. This works extremely well in practice to find models broadly compatible with the data. It is straightforward to incorporate this module into standard parameter estimation codes to do a direct comparison with cosmological data, and we will present such results in future work.

\begin{figure*}[]
\psfig{file=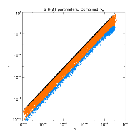,width=3.5in}\hfil
\psfig{file=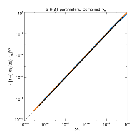,width=3.5in}
\caption{Slow-roll inflation (left) and DBI (right) consistency relations, combining the simulations with general and small $\gamma_0$ priors shown in Figs.~\ref{fig:2} and \ref{fig:3}, respectively. The left panel shows $r$ vs. $n_t$. In standard slow-roll inflation $r=-8n_t$ (dashed line), but as shown, DBI inflationary models depart significantly from this relation except in the case of slow roll DBI models (black points). In the right panel, we show $r \left[1+ \frac{108}{35} f_{NL} \right]^{1/2}$ vs. $-8 n_t$ describing the DBI consistency relation; all models satisfy this relation.}
\label{fig:4}
\end{figure*}

\begin{figure*}[]
\psfig{file=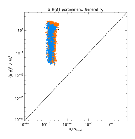,width=3.5in}\hfil
\psfig{file=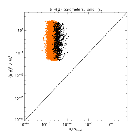,width=3.5in}
\caption{A consistency check of the field range bound (\ref{equ:dphi2}) for the general $\gamma_0$ simulation shown in Fig.~\ref{fig:2} (left) and the small $\gamma_0$ simulation shown in Fig.~\ref{fig:3} (right), with $\Delta \phi^2/M_P^2$ vs. the expression on the right hand side of Eq.~(\ref{equ:dphi2}), assuming ${\rm Vol}(X_5) = \pi^3$. The microscopic bound requires that $\Delta \phi^2/M_P^2$ be smaller than the right side of eq.~(\ref{equ:dphi2}), and {\it all} models in these simulations, which had a general $\epsilon_0$ prior applied, violate this bound. }
\label{fig:5}
\end{figure*}

\begin{figure*}[]
\psfig{file=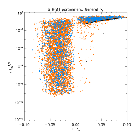,width=3.5in}\hfil
\psfig{file=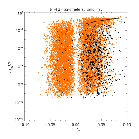,width=3.5in}
\caption{$r f_{NL}^{1/2}$ vs. $1-n_s$ for general $\gamma_0$ (left, Fig.~\ref{fig:2}) and small $\gamma_0$ (right, Fig.~\ref{fig:3}) simulations of DBI models. The black line is the lower limit on tensors for relativistic DBI models \cite{Lidsey}: $r > \frac{4(1-n_s)}{\sqrt{3 f_{NL}}}$. While this relation between $n_s$ and $r$ is satisfied by ultra-relativistic DBI models, this relation is not satisfied by the intermediate and slow roll limit DBI regimes.}
\label{fig:6}
\end{figure*}

\begin{figure*}[]
\psfig{file=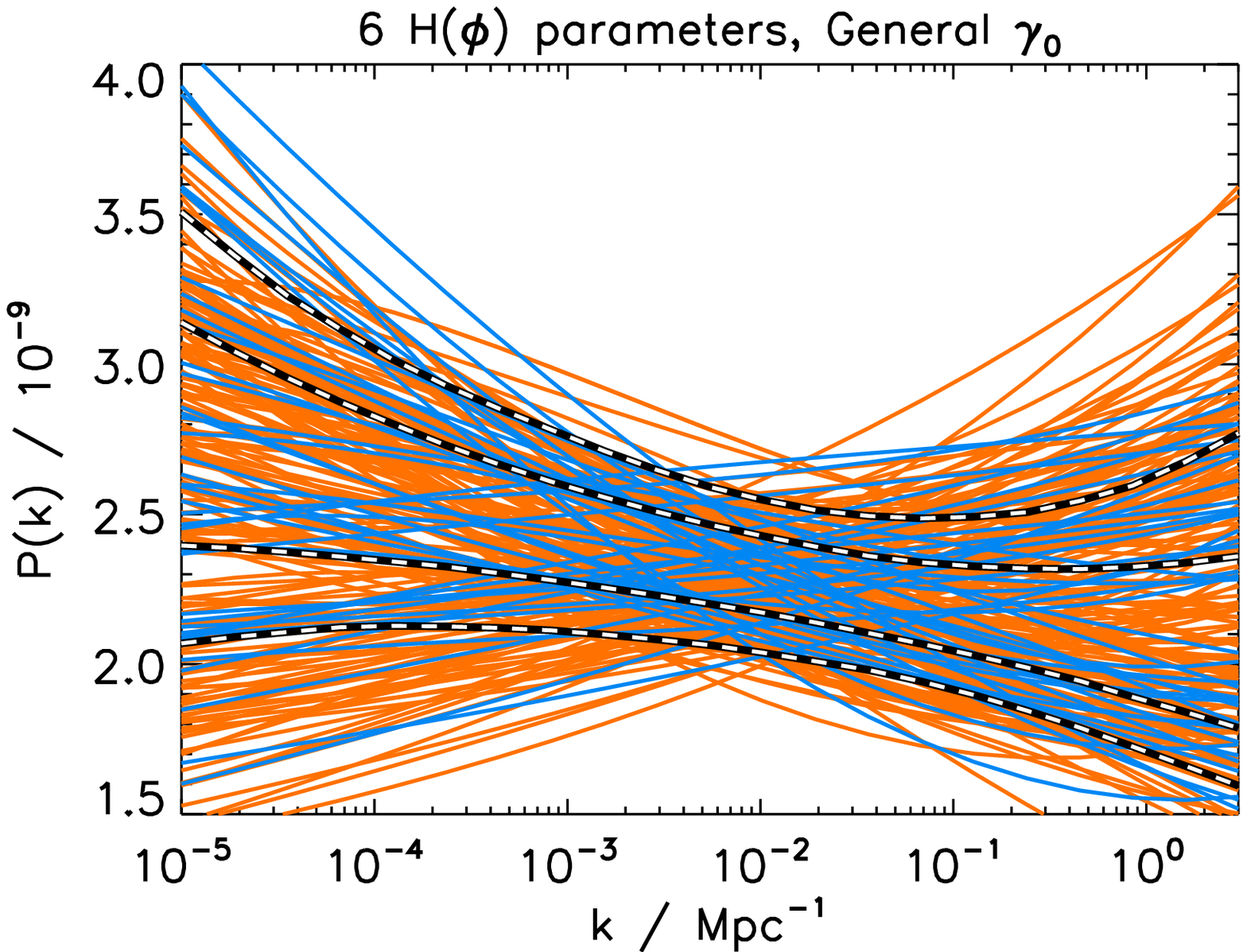,width=3.5in}\hfill
\psfig{file=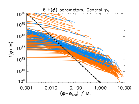,width=3.5in}\\
\psfig{file=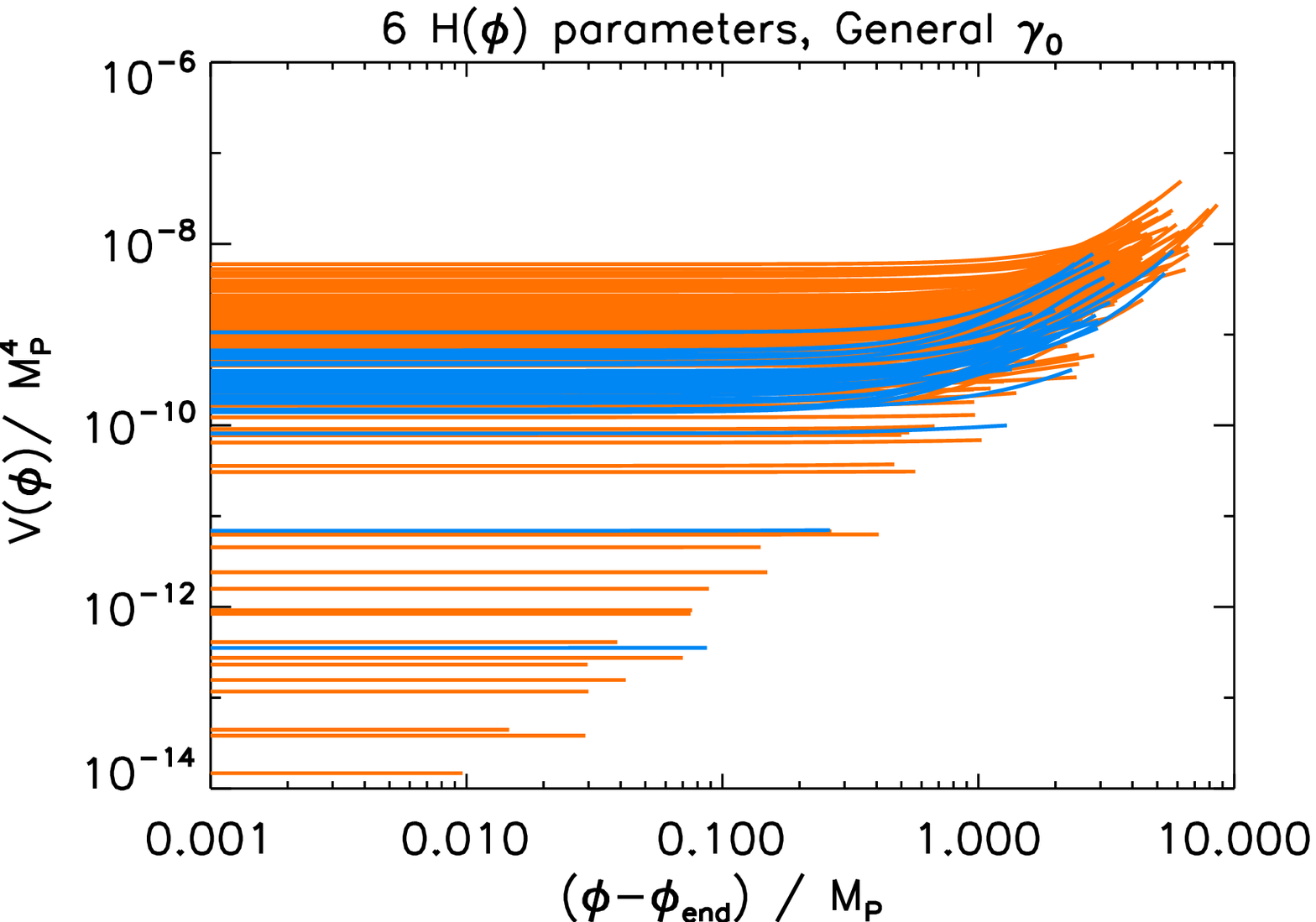,width=3.5in}\hfill
\psfig{file=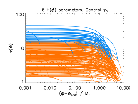,width=3.5in}
\caption{Input and output functions for the general $\gamma_0$ simulation from Fig.~\ref{fig:2}. From top-left in clockwise direction are: (a) $P_S(k)$, where the dashed lines are the 68\% and 95\% CL observational constraints from Ref.~\cite{Peiris:2006sj}, (b) $f(\phi)$, where the dashed line illustrates the shape of the AdS warp factor with $f(\phi) \propto (\phi - \phi_{\rm end})^{-4}$, (c) $V(\phi)$, and (d) $\gamma(\phi)$.}
\label{fig:7}
\end{figure*}

\begin{figure*}[]
\psfig{file=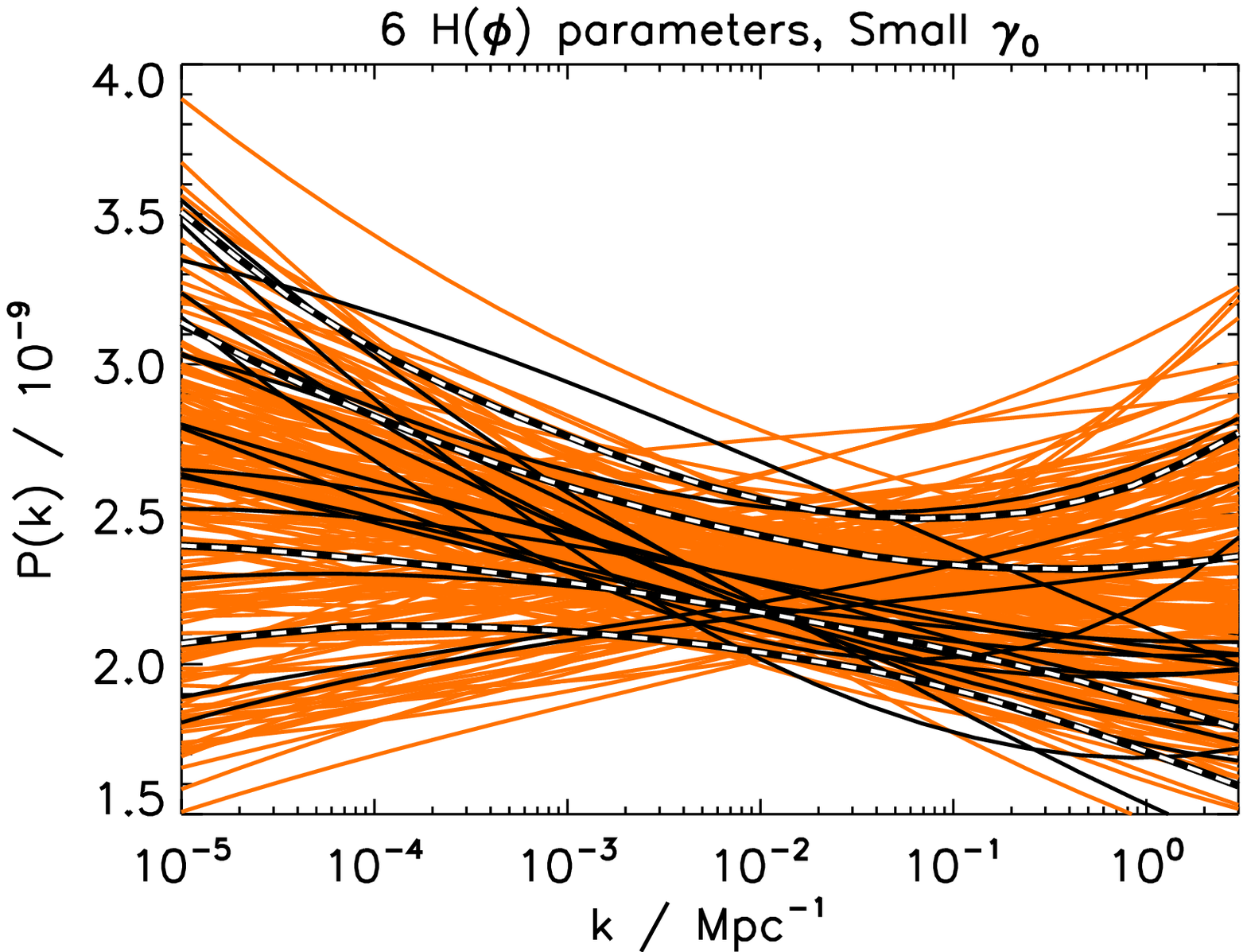,width=3.5in}\hfill
\psfig{file=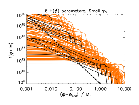,width=3.5in}\\
\psfig{file=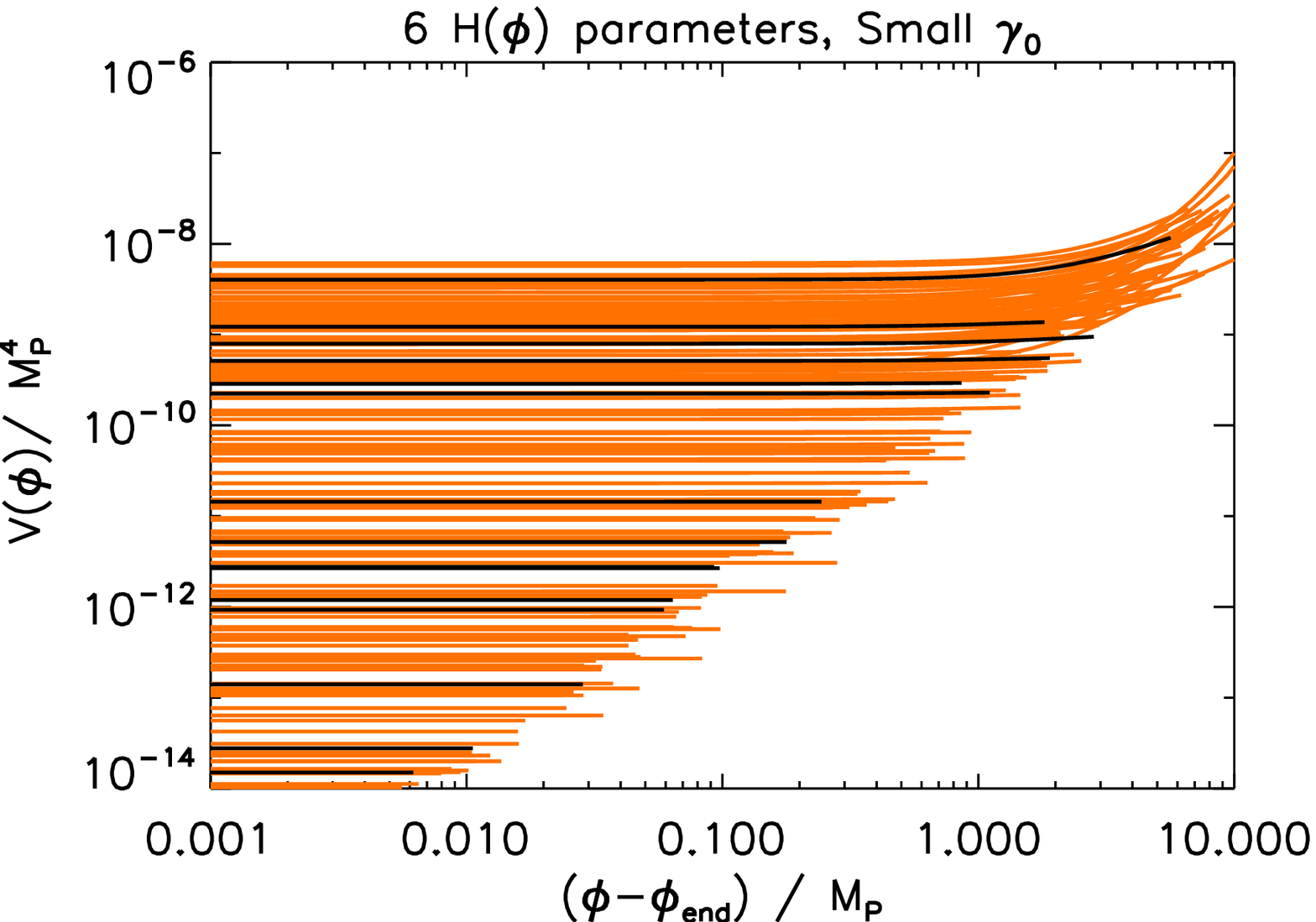,width=3.5in}\hfill
\psfig{file=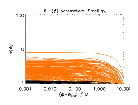,width=3.5in}
\caption{Same as Fig.~\ref{fig:7}, but for the small $\gamma_0$ simulation from Fig.~\ref{fig:3}.}
\label{fig:8}
\end{figure*}

\begin{figure*}[]
\psfig{file=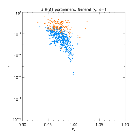,width=2.2in}\hfill
\psfig{file=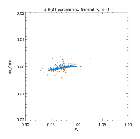,width=2.2in}\hfill
\psfig{file=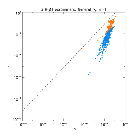,width=2.2in}\\
\psfig{file=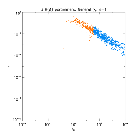,width=2.2in}\hfill
\psfig{file=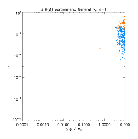,width=2.2in}\hfill
\psfig{file=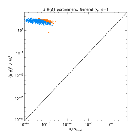,width=2.2in}
\caption{Observable parameters at $k_{\rm CMB}$ for a general $\gamma_0$ simulation with $M=2$ and $M'=1$ in equations~(\ref{eq:h}) and (\ref{eq:gamma}) respectively, showing models where inflation ends with $\epsilon \rightarrow 1$. The individual figures from top-left in clockwise direction are: (a) the scalar-to-tensor ratio $r$ vs. scalar spectral index $n_s$, (b) spectral index $n_s$ vs. running of the spectral index $\alpha_s=dn_s/d\ln k$, (c) $r$ vs. $n_t$, with a dashed line showing the standard slow roll expectation that $r=-8n_t$, (d)  $\Delta \phi^2/M_P^2$ vs. the expression on the right hand side of Eq.~(\ref{equ:dphi2}) assuming ${\rm Vol}(X_5) = \pi^3$, again showing that these models violate the microscopic bound on the field range, (e) $r$ vs. $\Delta \phi=\phi_0 -\phi_{\rm end}$, and (f) $r$ vs. non-Gaussianity parameter $f_{\rm NL}$, showing the limit from WMAP, $|f_{\rm NL}|<332$ \cite{fNL} with a vertical dashed line. The equivalent small $\gamma_0$ simulation gives qualitatively similar results.}
\label{fig:9}
\end{figure*}

\begin{figure*}[]
\psfig{file=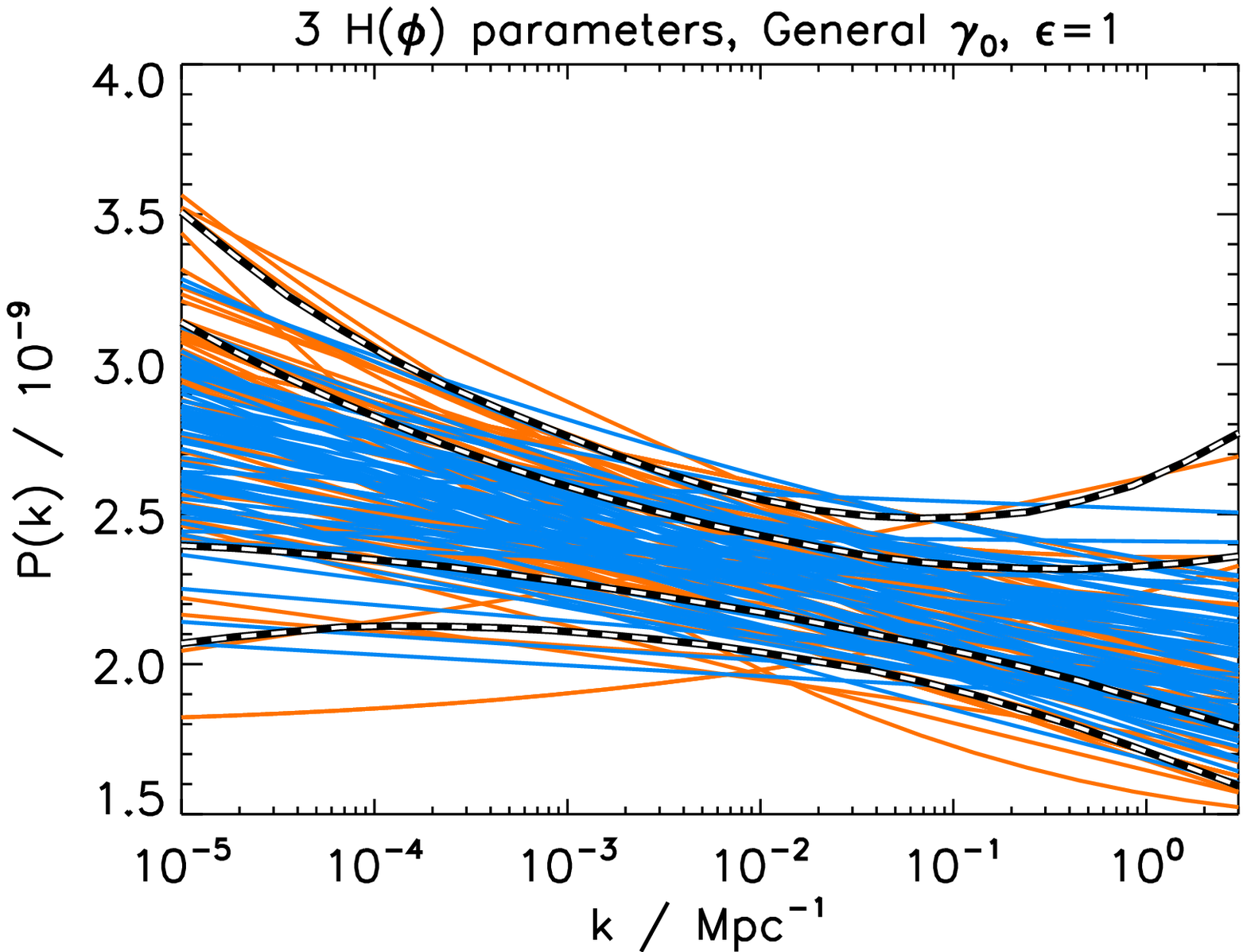,width=3.5in}\hfill
\psfig{file=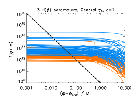,width=3.5in}\\
\psfig{file=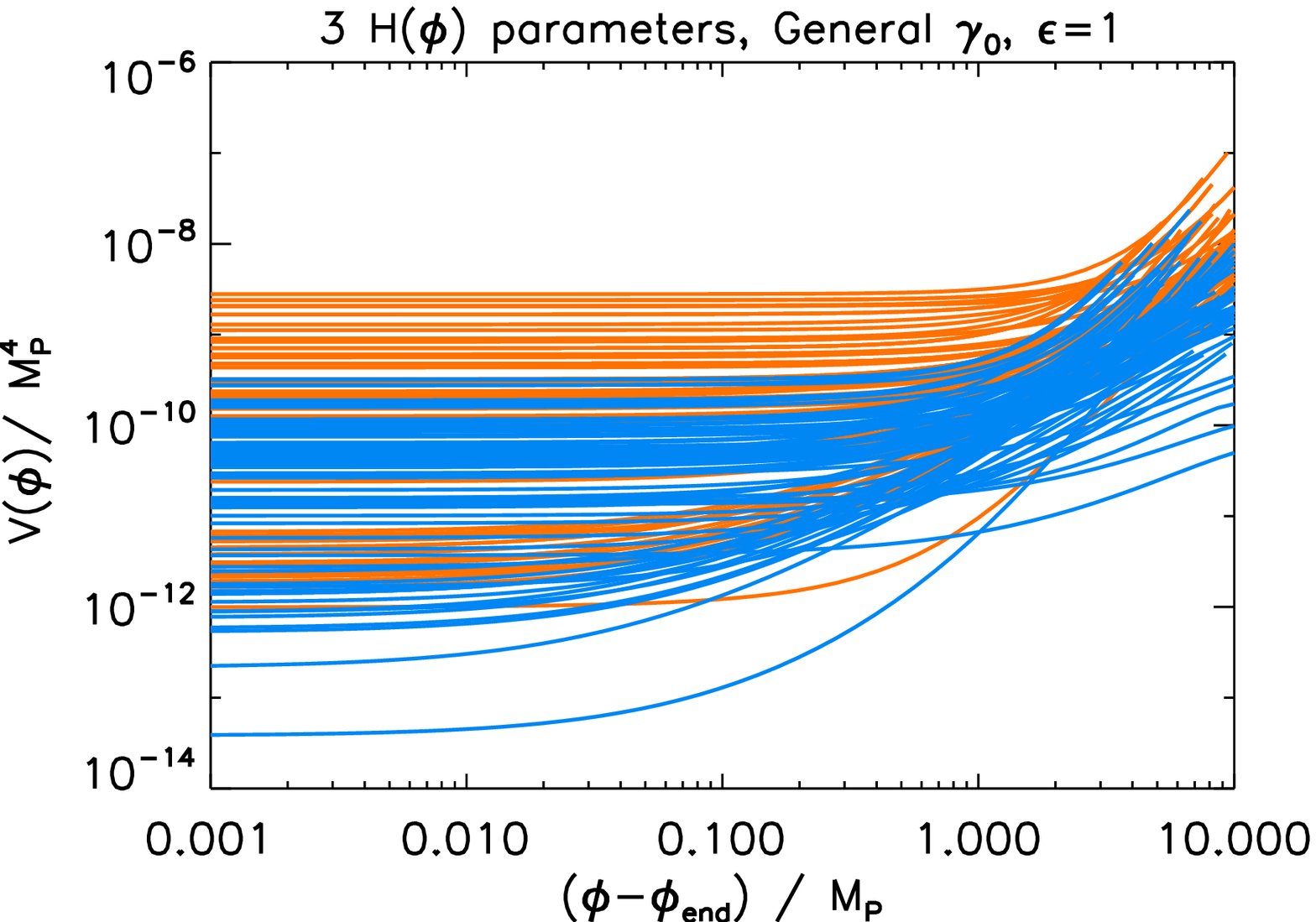,width=3.5in}\hfill
\psfig{file=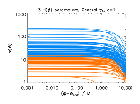,width=3.5in}
\caption{Input and output functions for the models in Fig.~\ref{fig:9}. The panel descriptions are the same as those of Fig.~\ref{fig:7}. The equivalent small $\gamma_0$ simulation gives qualitatively similar results. }
\label{fig:10}
\end{figure*}

\begin{figure*}[]
\psfig{file=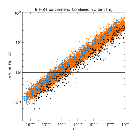,width=3.3in} \hfill
\psfig{file=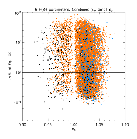,width=3.3in} \\
\caption{(a) Tensor/scalar ratio $r$ (left) and (b) scalar spectal index $n_s$ (right) vs. the microphysical constraint on the left hand side of Eq.~(\ref{equ:Fig11}) assuming ${\rm Vol}(X_5) = \pi^3$, combining general and small $\gamma_0$ priors with $M=5$ and $M'=1$ in equations~(\ref{eq:h}) and (\ref{eq:gamma}). Unlike previous simulations shown, these models have a small $\epsilon_0$ prior applied. The points below the horizontal line are DBI inflationary models allowed by the microphysical bound while models above this line violate the bound and are unphysical. The shapes of input and output functions for models satisfying this bound have qualitatively the same properties as shown before. 
From (a) we see that consistent intermediate and ultra-relativistic DBI models satisfy a very strong upper bound on tensors: $r < 10^{-15}$. This limit is consistent with the bound (\ref{equ:LidseyBound}) since the second plot in Fig.~\ref{fig:FinalPlot} shows that all consistent models with large $f_{NL}$ (so that (\ref{equ:LidseyBound}) applies) have a blue spectrum $n_s >1$.
}
\label{fig:FinalPlot}
\end{figure*}

\section{Results}
\label{sec:results}

In Figures \ref{fig:KinneyPlot} to \ref{fig:FinalPlot}, we present our results for observables at CMB scales as well as the functions associated with the underlying models. We classify DBI models by their value of $\gamma$ at CMB scales: $1< \gamma_{\rm CMB} < 1.1$, $1.1< \gamma_{\rm CMB} < 10$, $\gamma_{\rm CMB}>10$ define the {\sf slow roll} (black), {\sf intermediate DBI} (orange) and {\sf ultra-relativistic DBI} (blue) regimes, respectively.

Different combinations of the priors on $\gamma_0$ and $\epsilon_0$ described in \S \ref{subsec:algorithm} have been applied to each figure. All figures except Fig.~\ref{fig:FinalPlot} use the general $\epsilon_0$ prior; the latter has the small $\epsilon_0$ prior applied. As specified in the figure captions, sometimes we will show the general and small $\gamma_0$ prior simulations separately to emphasize different points of interest, and sometimes we will combine these simulations. While in general the plots with combined priors have different \emph{measures} on models in each simulation, this does not matter for our purpose, which is to investigate the functional parameters and predictions of these models qualitatively.

In addition, all figures except Figs.~\ref{fig:9} and \ref{fig:10} show simulations run with $M=5$ and $M'=1$ in equations~(\ref{eq:h}) and (\ref{eq:gamma}); the latter used $M=2$ and $M'=1$. In the case with six $H(\phi)$ parameters, we found that all of the models had inflation ending with tachyonic transitions. This is because models with significant higher order derivatives of $H(\phi)$ which end inflation with $\epsilon=1$ have a strong tendency to provide insufficient $e$-folds of inflation to satisfy our requirements, and are therefore rejected from our simulations.  In order to get a fraction of models where inflation ended with the end of slow roll, $\epsilon=1$, we had to reduce the number of $H(\phi)$ parameters to three; even in this case, tachyonic transitions were the dominant mechanism for the end of inflation. In Figs.~\ref{fig:9} and \ref{fig:10} we show the properties of only the models where inflation ended with $\epsilon=1$.\\

Here we highlight and comment on what we consider to be the most important results of this study.
Further details can be found in the figures and the figure captions.

Fig.~\ref{fig:KinneyPlot} compares the distributions of points in the $n_s$ vs $r$ plane from a Kinney-style single-field slow roll simulation \cite{Kinney} and a DBI simulation. Note that the algorithm used to create the slow roll simulation is significantly different from ours; among other dissimilarities, it makes use of both forward and backward integration of the single-field slow roll flow equations, while our algorithm only uses forward integration of the generalized flow equations. The slow roll simulation is shown here because it adopts a similar philosophy that ``fundamental'' quantities of interest which are not strictly fixed by theory, such as the potential, should be Monte-Carlo'd rather than fixed to specific functions when investigating the range of observable properties predicted by a given model. More detailed inferences should not be made from this comparison because of the differences in algorithms.

A paucity of points is apparent in the DBI simulation as $|n_s-1| \to 0$. We explain this as follows: 
For perfect AdS warp factors $f(\phi) \propto \left(\phi - \phi_{\rm end}\right)^{-4}$ and a linear Hubble parameter $H(\phi) \propto  \left(\phi-\phi_{\rm end}\right)$ one may show analytically that $n_s = 1$ to all orders in inflationary slow roll parameters \cite{Tye}. However, it has also been shown that non-linear corrections to $H(\phi)$ and deviations of the warp factors $f(\phi)$ from the AdS limit lead to deviations from scale invariance \cite{Shiu}. This is what we are seeing here. Our warp factors are generally flatter than AdS for the regime of cosmological interest, and $H''(\phi)$ is often non-negligible. Below we explain why we believe our algorithm preferentially selects non-AdS warp factors. The same argument then explains the lack of models leading to perfect scale invariance.

Fig.~\ref{fig:2} and \ref{fig:3} present the results for the basic cosmological observables evaluated on CMB scales for {\sf general $\gamma_0$ prior} and {\sf small $\gamma_0$ prior}, respectively.
As illustrated in Fig.~\ref{fig:2}(a), which shows $r$ vs. $n_s$, models with detectably large values for the non-Gaussianity parameter $f_{NL}$ satisfy a {\it lower} limit on the tensor-to-scalar ratio $r$ if the scalar spectrum is red $n_s < 1$. This bound was derived in \cite{Lidsey_Seery, Lidsey} and may be written as
\beq
\label{equ:LidseyBound}
r > \frac{4}{\sqrt{3 f_{NL}}} (1-n_s)\, .
\eeq
Notice that the bound (\ref{equ:LidseyBound}) only applies if $f_{NL} > 1$, as can be seen from the $r$ vs. $n_s$ plot in Fig.~\ref{fig:3}(a). Since current CMB and large-scale structure data are strongly indicating a red scalar spectrum \cite{Observations}, Fig.~\ref{fig:2} then implies the exciting prediction that DBI models with detectably large non-Gaussianities correlate with detectably large tensors. Unfortunately, this conclusion ignores an important theoretical consistency constraint that should be imposed on any viable model. As Ref.~\cite{BauMcA} showed, the inflation field range in warped D-brane inflation is geometrically limited, and in explicit models can be written in the following simple form
 \beq
 \frac{\Delta \phi}{M_P} < \frac{2}{\sqrt{N}}\, ,
 \eeq
where $N \gg 1$. By the Lyth bound \cite{LythBound} this impossibility of super-Planckian field variation during inflation implies a strong {\it upper}  limit on $r$. For the generalized warp factors studied in this paper the field range bound is written in the generalized form (\ref{equ:dphi2}).
Fig.~\ref{fig:5} illustrates this bound for the models shown in Fig.~\ref{fig:2} and \ref{fig:3}. The plots show rather dramatically that all models in Fig.~\ref{fig:2} and \ref{fig:3} violate the microscopic bound of \cite{BauMcA}.

To find models that satisfy the field range bound (\ref{equ:dphi2}) we ran separate simulations with {\sf small $\epsilon_0$ prior}. The results are illustrated in Fig.~\ref{fig:FinalPlot}.  
Here, the $y$--axis represents the compactification constraint on the warped throat volume (\ref{equ:PlanckMass}) written in the following form
\beq
\label{equ:Fig11}
\frac{{\rm Vol}(X_5)}{\pi} \int_{\phi_{\rm end}}^{\phi_0} \left(\frac{\d \phi}{M_P} \right)  \left(\frac{\phi}{M_P} \right)^5 (f(\phi) M_P^4) < 1\, . 
\eeq
This condition has to be imposed as a consistency constraint on all models.  
Models that violate (\ref{equ:Fig11}) correspond to a mismatch between the field range required by the specific inflationary model and the field range allowed by the compactification.
We find that (\ref{equ:Fig11}) can only be satisfied if the field range from IR ($\phi_{\rm end}$) to UV ($\phi_0$) is very small (in Planck units). This is because the magnitude of $f$ is constrained to be large by the normalization of the scalar spectrum, so the size of the integral can't be made small by making $f$ small. Instead the integral can only be small if the range of integration is small. By the Lyth bound this small range for the field variation corresponds to a very small tensor signal.

From Fig.~\ref{fig:FinalPlot} we see that theoretically consistent intermediate and ultra-relativistic DBI models satisfy a very strong upper bound on tensors: $r < 10^{-15}$.
This limit is consistent with the bound (\ref{equ:LidseyBound}) since the second plot in Fig.~\ref{fig:FinalPlot} shows that all consistent models with large $f_{NL}$ (so that (\ref{equ:LidseyBound}) applies) have a blue spectrum $n_s >1$. These models could therefore be falsified in the near future if the current indications of a red spectrum gains in statistical significance to be more than $3\sigma$.

We emphasize, however, that the bound of $r < 10^{-15}$ is not absolute and in particular does {\it not} apply to slow roll models with small $f_{NL}$. The upper limit on $r$ for consistent slow roll models (black points) in Fig.~\ref{fig:FinalPlot} depends on the lower limit of the prior on $\gamma_0$. For very small $\gamma$, larger $r$ becomes consistent with the field range bound.
(Similar conclusions were obtained by Bean {\it et al.} \cite{Bean}. In particular, they showed that only extreme slow models with $\gamma_{\rm CMB} -1 < 10^{-7}$ survive the bound of \cite{BauMcA}).
However, the theoretical considerations of \cite{BauMcA} still apply, which predict an ultimate upper limit on $r$ for slow roll models. This limit is far below the detection sensitivity of any realistic future experiments.

We now comment on the shapes of the empirical $f(\phi)$ derived from our simulations as compared to the analytic AdS warp factor $f(\phi) \propto \phi^{-4}$ studied in the literature in connection with DBI models. In our simulations with 6 $H(\phi)$ parameters, many of the recovered $f(\phi)$, as seen in Figs \ref{fig:7}(b) and \ref{fig:8}(b), exhibit power-law behavior similar to the AdS warp factor, but are somewhat shallower than the AdS case. Further, there are some models which start out with power-law behavior and flatten out towards the end of inflation. In the case of our simulations with 3 $H(\phi)$ parameters, where we consider models where inflation ends with $\epsilon=1$, the derived $f(\phi)$ shapes are exclusively of the latter kind. Our work shows that, qualitatively, there are some differences between the shapes of warp factors found through the empirical process described in our algorithm and the analytical considerations so far. However, given the large parameter space explored in our simulations, there was no \emph{a priori} expectation that we would recover this particular function. Therefore, it is encouraging that the derived functions are not, in fact, dramatically different from the analytic form. Given that our prior is broad enough to allow $\phi^{-4}$ warp factors to be generated in a Monte-Carlo fashion, two possibilities are that either the shallower functional forms are more likely to be encountered in the prior, or that the penalty function used in our Metropolis-Hastings algorithm prefers shallower functions over steeper ones. In a future study, we plan to make a full investigation of the likelihood of the analytic AdS warp factor compared to the derived empirical warp factors, in order to test whether the data in fact prefers shallower functions for $f(\phi)$.

\vfil

\section{Conclusion}
\label{sec:conclusion}

In this paper we studied the basic phenomenology of D-brane inflation models with general speed of sound. We developed a general Monte-Carlo formalism for studying models with generalized warp factors and inflaton potentials.  Most models show large deviations of the speed of sound from the speed of light resulting in exciting observational signatures. Non-Gaussianities are typically large enough to be observable and the standard slow roll consistency relation is violated at a non-negligible level.
However, we also showed that most of these phenomenological models cannot be embedded into a consistent string compactification. In particular, the vast majority of models violates the field range bound of \cite{BauMcA}. 
UV models of DBI inflation which obey the field range bound are of two distinct types:\footnote{In addition, there is the 'IR model' \cite{ChenIR} which requires a separate analysis.}
\begin{enumerate}
\item Slow roll models with unobservably small $f_{NL}$.
\item Relativistic DBI models with observable $f_{NL}$, but blue scalar spectrum, $n_s > 1$.
\end{enumerate}
Let us therefore imagine that future observations yield a firm measurement of a red scalar spectrum, {\it i.e.} a statistically significant detection of $n_s < 1$ after marginalizing over (or detecting) $r$ and $dn_s/d\ln k$ and the ``late-time'' cosmological parameters. What would this imply for brane inflation in general and the DBI limit in particular? As indicated above, such an observation of the scalar spectrum would effectively rule out relativistic DBI models, without having to perform any measurements of non-Gaussianities. This illustrates that cosmological observations are quickly becoming precise enough, so that many theoretical ideas are highly constrained. DBI inflation is highly falsifiable and might in fact very soon be ruled out by the data.\footnote{To be precise, we emphasize that this statement at present only applies to realizations of the DBI mechanism using D3-branes on Calabi-Yau cones.
In this sense our conclusions do {\it not} represent model-independent constraints on the DBI mechanism.  In particular, all our statements are restricted to the concrete realizations of DBI inflation that we described in \S\ref{sec:background}.
Given that the strong constraints on these models are mostly driven by the geometrical field range bound of \cite{BauMcA}, it would be very interesting to find generalizations of this class of models that evade the bound.  Concrete efforts to construct such scenarios are under way \cite{DBIv2Eva, DBIv2SarahLouis}.}
This conclusion motivates considering slow roll models of brane inflation. However, these models suffer from a version of the supergravity eta-problem \cite{KKLMMT}; moduli stabilization effects typically induces large corrections to the inflation mass and inflation cannot occur. Given the prospect that relativistic DBI inflation via D3-brane motion on Calabi-Yau cones might soon be ruled out by the data, it becomes important to understand whether the eta-problem for slow roll brane inflation can be overcome \cite{BDKM, Burgess, Krause}. 

We are living in the fortunate age where the cosmological data are becoming precise enough to significantly constrain theoretical models of the early universe.

\addtocontents{toc}{\SkipTocEntry}
\section*{Acknowledgments}
We are grateful to Rachel Bean, 
Igor Klebanov, Liam McAllister, Daniel Mortlock, Sarah Shandera, Eva Silverstein, and Henry Tye for helpful discussions. 
We thank Richard Easther and Liam McAllister for comments on a draft.
HVP is supported by NASA through Hubble Fellowship grant \#HF-01177.01-A awarded by the Space Telescope Science Institute, which is operated by the Association of Universities for Research in Astronomy, Inc., for NASA, under contract NAS 5-26555. HVP acknowledges the hospitality of the TAPIR group at Caltech, where part of this work was carried out as a Moore Program visitor. AC acknowledges support from NSF CAREER AST-0645427.


\newpage
\appendix 
\section{Monte-Carlo Algorithm for DBI Inflation}
\label{appendix:algorithm}

In \S\ref{subsec:algorithm}, we gave a broad outline of the Monte-Carlo algorithm that we use to produce the numerical results of \S\ref{sec:results}. In this Appendix, we describe its implementation in detail.

The sign convention is tied as follows to the physical picture of a brane evolving from the UV mouth of the warped throat towards its IR tip. As time increases, the brane falls into the throat towards the end of inflation. We will call this direction \emph{forward}. ${\rm d}\phi < 0$ as the brane moves forward. $N_e=0$ at the end of inflation by convention, and large at $\phi=0$; {\it i.e.} $N_0>N_{\rm CMB}>N_{\rm end}$ where $N_{\rm end}=0$. In practice, we set $N_0$ to some large value, and once we find $N_{\rm end}$ by evolving forward, we just translate the zero point so that $N_{\rm end}=0$. 

\addtocontents{toc}{\SkipTocEntry}
\subsection{Initial Conditions} \label{sssec:ic}
First, since we define $\phi=0$ be the UV end of the throat, then by definition,
\beq
h_{\rm UV}^{-1} = h^{-1}(0) \equiv 1\, ,
\eeq
and
\beq
f_0^{-1} = f^{-1}(0) = T_3\, .
\eeq

As stated in the main text, we apply several different types of priors on $\gamma_0$ and $\epsilon_0$, as follows:
\begin{itemize}
\item  {\sf General $\gamma_0$ initial conditions}: We draw $\log_{10} \gamma_0$ randomly from the broad flat prior $ [0,1]$.
\item {\sf Small $\gamma_0$ initial conditions}: We draw $\log_{10} \gamma_0$ randomly from the narrow flat prior $ [0,10^{-5}]$.
\item {\sf General $\epsilon_0$ initial conditions}: We draw $\log_{10} \epsilon_0$ randomly from the broad flat prior $[-10,0]$. 
\item {\sf Small $\epsilon_0$ initial conditions}: We draw $\log_{10} \epsilon_0$ randomly from the narrow flat prior $[-21,-9]$. 
\end{itemize}
The rest of the algorithm is identical for all priors.

We draw the initial amplitude of the power spectrum of scalar density perturbations at 
$\phi = 0$
and its value at the CMB scale (taken to be $k_{\rm CMB}=0.02$ Mpc$^{-1}$) from flat priors,
\begin{eqnarray}
P_0 = P_s(\phi=0)&\in& [5\times10^{-10},1\times10^{-8}] \nonumber \\  
P_s(k_{\rm CMB}) &\in& [5\times10^{-10},1\times10^{-8}]\, .
\end{eqnarray}

We select a limited number of the slow variation parameters in equations~(\ref{eq:h}, \ref{eq:gamma})
and Monte-Carlo over their initial conditions using the following flat priors.
\begin{eqnarray}
\eta_0 &\in&[-0.05,0.1] \nonumber \\
\xi_0 &\in&[-5\times10^{-3},5\times10^{-3}] \nonumber \\
^3\lambda_0 &\in&[-5\times10^{-4},5\times10^{-4}] \nonumber \\
^4\lambda_0 &\in&[-5\times10^{-5},5\times10^{-5}] \nonumber \\
^5\lambda_0 &\in&[-5\times10^{-6},5\times10^{-6}] \nonumber \\
\kappa_0 &\in&[-0.1,0] \nonumber \\
^1 \alpha_{0} &\in&[-0.01,0.01].
\label{eqn:params}
\end{eqnarray}
Other (higher order) parameters are set to zero at the initial conditions, and therefore remain zero for the entire evolution. 

With the choice of parameters in Eq.~(\ref{eqn:params}), we have expanded $H(\phi)$ and $\gamma(\phi)$ polynomials to 6th- and 2nd-order in the scalar field $\phi$, respectively. Such an expansion is more than adequate to capture, for example, the behavior of functions such as $f(\phi) \propto (\mu +  \phi)^{-4}$ through Eq.~(\ref{equ:g}). We have checked that our qualitative results do not change by adding higher order slow evolution parameters to the dynamics.\\

\addtocontents{toc}{\SkipTocEntry}
\subsection{Evolution and Matching to CMB Scales} \label{sssec:cmbmatch}

We numerically evolve the flow equations forward into the throat ($\phi$ decreases) until the matching condition 
\beq
\frac{\gamma_{\rm CMB}}{8 \pi^2 \epsilon_{\rm CMB}} \left( \frac{H_{\rm CMB}}{M_P} \right)^2 = P_s(k_{\rm CMB}) \label{eq:cmbmatch}
\eeq
is satisfied. Then we need to identify the $\phi$ value corresponding to $k_{\rm CMB}$. Note that unlike in slow roll inflation, the horizon exit for scalar and tensor modes for a given physical scale $k_{\rm CMB}$ happens at \emph{different} $\phi$ values, $\phi_s^{\rm CMB}$ and  $\phi_t^{\rm CMB}$ respectively. These are related by
\beq
a(\phi_s)H(\phi_s)\gamma(\phi_s)|_{\rm CMB} = a(\phi_t)H(\phi_t)|_{\rm CMB}.
\eeq
Rewriting this in a way to minimize numerical errors in the matching,
\begin{eqnarray}
&& N_e^0 - N_e^s(\phi_s) + \ln [H(\phi_s)\gamma(\phi_s)] \nonumber \\
&=&  N_e^0 - N_e^t(\phi_t)   +  \ln H(\phi_t), 
\end{eqnarray}
where $N_e^0$ is the \emph{arbitrary} number of $e$-folds associated with $\phi=0$.

Models which never satisfy the matching condition (\ref{eq:cmbmatch}) are rejected. In addition, throughout the evolution, we impose the following constraints: 
\begin{enumerate}
\item In order to be compatible with the physical picture of single throat UV DBI inflation, the model must have a monotonically increasing warp factor, {\it i.e.} $d f/d\phi > 0$ must be satisfied at all times during the evolution.
\item $k(\phi)$ must be a monotonic function. This requires $\epsilon + \kappa < 1$, where
\beq
\epsilon + \kappa = \sqrt{\frac{2 \epsilon}{\gamma}} \frac{d
\ln (\gamma H)}{d \frac{\phi}{M_P}}\, .
\eeq
Since $\sqrt{\epsilon/\gamma} < 1$, the condition $\epsilon + \kappa < 1$ implies
\beq
\frac{d \ln (c_s H^{-1})}{d \frac{\phi}{M_P}} < 1\, .
\eeq
In other words, the sound horizon shouldn't grow exponentially as the field evolves over Planckian distances.  
\item In the single throat scenarios that we are modeling, $\kappa(\phi)$ is constrained to be negative.
To see this note that
\beq
\kappa(\phi) = - \frac{d \ln c_s}{d N_e} = \frac{d \ln \gamma}{d N_e}\, .
\eeq
For a brane moving into a throat
$\gamma = [1- f \dot \phi^2]^{-1/2}$ is necessarily monotonically increasing, because both $f(\phi)$ and $\dot \phi$ are monotonically increasing as the brane speeds up while moving towards larger warping.
With our convention $\d N_e = - H \d t$ this implies that $\kappa(\phi) < 0$.
\item $\gamma(\phi) \geq 1$, by definition.
\item Successful models must be able to accommodate {\it at least} the range of physical scales $k \in [1\times10^{-5}, 3]$ Mpc$^{-1}$ between $\phi=0$ and $\phi_{\rm end}$.
\end{enumerate}
Models failing any one of these requirements are rejected from the simulation.

\addtocontents{toc}{\SkipTocEntry}
\subsection{End of Inflation} \label{sssec:infend}

We consider two distinct scenarios for the end of inflation: tachyonic instability and the end of slow roll.
In order to capture the uncertainties in the reheating energy scale which translate into significant uncertainties in the number of $e$-folds of inflation between the CMB scale exiting the horizon and the end of inflation \cite{Kinney:2005in} we draw a random number of $e$-folds to be sampled from 
$N_e^\star  \in[40,70]$. Then we evolve forward towards the end of inflation so that $N_{\rm CMB}-N_{\rm end}=N_e^\star$. Defining $\phi = \phi_{\rm end}$ and computing $f_{\rm end} = f(\phi_{\rm end})$ and $h_{\rm end} = T_3 f_{\rm end}$, we check that $h_{\rm end}^{-1} \in [10^{-10},10^{-2}]$.
 This is our model for inflation ending via a tachyonic transition. If slow roll ends ({\it i.e.}~$\epsilon \to 1$) before $N_e^\star$ is reached, we identify this as the end of inflation if this occurs at least 40 $e$-folds from the CMB scale, taking the number of $e$-folds between $\phi_{\rm CMB}$ and $\epsilon = 1$ as $N_e^\star$. If neither of these conditions are satisfied, the model is rejected.


\newpage
\begin{thebibliography}{99}
\frenchspacing

\bibitem{Inflation}
  A.~H.~Guth,
  Phys.\ Rev.\ D {\bf 23}, 347 (1981);

  A.~D.~Linde,
  Phys.\ Lett.\ B {\bf 108}, 389 (1982);

  A.~Albrecht and P.~J.~Steinhardt,
  Phys.\ Rev.\ Lett.\  {\bf 48}, 1220 (1982).

\bibitem{Dvali}
  G.~R.~Dvali and S.~H.~H.~Tye,
  Phys.\ Lett.\ B {\bf 450}, 72 (1999).
  
\bibitem{KKLMMT}
  S.~Kachru, R.~Kallosh, A.~Linde, J.~Maldacena, L.~McAllister and S.~P.~Trivedi,
  JCAP {\bf 0310}, 013 (2003).

\bibitem{otherbraneinflation}
S.~Alexander, 
Phys. Rev. D {\bf 65}, 023507 (2002); 
G.~Dvali, Q.~Shafi and S.~Solganik, 
[arXiv:hep-th/0105203];
C.~P.~Burgess, M.~Majumdar, D.~Nolte, F.~Quevedo, G.~Rajesh and
R.~J.~Zhang, 
JHEP
{\bf 07}, 047 (2001);
 J.~H.~Brodie and D.~A.~Easson,
  JCAP {\bf 0312}, 004 (2003);
  K.~Becker, M.~Becker and A.~Krause,
  Nucl.\ Phys.\  B {\bf 715}, 349 (2005).

\bibitem{DBI}
  E.~Silverstein and D.~Tong,
  Phys.\ Rev.\ D {\bf 70}, 103505 (2004);
    M.~Alishahiha, E.~Silverstein and D.~Tong,
  Phys.\ Rev.\ D {\bf 70}, 123505 (2004).

\bibitem{Kahler}
  J.~P.~Conlon and F.~Quevedo,
  JHEP {\bf 0601}, 146 (2006);
  J.~R.~Bond, L.~Kofman, S.~Prokushkin and P.~M.~Vaudrevange,
  arXiv:hep-th/0612197.

\bibitem{racetrack}
  J.~J.~Blanco-Pillado {\it et al.},
  JHEP {\bf 0609}, 002 (2006).
  
\bibitem{Nflation}
  S.~Dimopoulos, S.~Kachru, J.~McGreevy and J.~G.~Wacker,
  arXiv:hep-th/0507205;
   R.~Easther and L.~McAllister,
  JCAP {\bf 0605}, 018 (2006).

\bibitem{Observations}
  D.~N.~Spergel {\it et al.}  [WMAP Collaboration],
 Astrophys.\ J.\ Suppl.\  {\bf 148}, 175 (2003);

  H.~V.~Peiris {\it et al.},
  Astrophys.\ J.\ Suppl.\  {\bf 148}, 213 (2003);

  D.~N.~Spergel {\it et al.},
  arXiv:astro-ph/0603449;

 M.~Tegmark {\it et al.},
  Phys.\ Rev.\  D {\bf 74}, 123507 (2006);
  
  S.~Cole {\it et al.}  [The 2dFGRS Collaboration],
  Mon.\ Not.\ Roy.\ Astron.\ Soc.\  {\bf 362}, 505 (2005).
  
\bibitem{Chen}
  X.~Chen, M.~x.~Huang, S.~Kachru and G.~Shiu,
  JCAP {\bf 0701}, 002 (2007).
  
\bibitem{Lidsey_Seery}
  J.~E.~Lidsey and D.~Seery,
   Phys.\ Rev.\  D {\bf 75}, 043505 (2007).

\bibitem{BauMcA}
  D.~Baumann and L.~McAllister,
  Phys.\ Rev.\ D {\bf 75}, 123508 (2007).

\bibitem{Bean}
  R.~Bean, S.~E.~Shandera, S.~H.~H.~Tye and J.~Xu,
  arXiv:hep-th/0702107.

\bibitem{Lidsey}
  J.~E.~Lidsey and I.~Huston,
  arXiv:0705.0240 [hep-th].

\bibitem{KKLT}
  S.~Kachru, R.~Kallosh, A.~Linde and S.~P.~Trivedi,
  Phys.\ Rev.\ D {\bf 68}, 046005 (2003).

\bibitem{FluxReview}
  M.~R.~Douglas and S.~Kachru,
  arXiv:hep-th/0610102.

\bibitem{GKP}
  S.~B.~Giddings, S.~Kachru and J.~Polchinski,
  Phys.\ Rev.\ D {\bf 66}, 106006 (2002).

\bibitem{RS}
  L.~Randall and R.~Sundrum,
  Phys.\ Rev.\ Lett.\  {\bf 83}, 3370 (1999).

\bibitem{ChenIR}
  X.~Chen,
  JHEP {\bf 0508}, 045 (2005)
  [arXiv:hep-th/0501184];
  X.~Chen,
  Phys.\ Rev.\ D {\bf 72}, 123518 (2005)
  [arXiv:astro-ph/0507053].

\bibitem{BDKMMM}
  D.~Baumann, A.~Dymarsky, I.~R.~Klebanov, J.~Maldacena, L.~McAllister and A.~Murugan,
   JHEP {\bf 0611}, 031 (2006).

\bibitem{Tye}
  S.~E.~Shandera and S.~H.~Tye,
  JCAP {\bf 0605}, 007 (2006).

\bibitem{BDKM}
  D.~Baumann, A.~Dymarsky, I.~R.~Klebanov and L.~McAllister,
  arXiv:0706.0360 [hep-th];
   D.~Baumann, A.~Dymarsky, I.~R.~Klebanov, L.~McAllister, and P.~Steinhardt
 arXiv:0705.3837 [hep-th].
  
\bibitem{Garriga}
  J.~Garriga and V.~F.~Mukhanov,
  Phys.\ Lett.\  B {\bf 458}, 219 (1999).

\bibitem{ClineReview}
  J.~M.~Cline,
  arXiv:hep-th/0612129.
  
\bibitem{Juan}
  J.~M.~Maldacena,
  JHEP {\bf 0305}, 013 (2003).
  
\bibitem{fNL}
  P.~Creminelli, L.~Senatore, M.~Zaldarriaga and M.~Tegmark,
  JCAP {\bf 0703}, 005 (2007).

\bibitem{MZ06} 
K.~M.~Smith and M.~Zaldarriaga,
  arXiv:astro-ph/0612571.

\bibitem{Smith:06}
  T.~L.~Smith, H.~V.~Peiris and A.~Cooray,
  Phys.\ Rev.\  D {\bf 73}, 123503 (2006).

\bibitem{KS}
  I.~R.~Klebanov and M.~J.~Strassler,
  JHEP {\bf 0008}, 052 (2000).

\bibitem{LythBound}
  D.~H.~Lyth,
  Phys.\ Rev.\ Lett.\  {\bf 78}, 1861 (1997).
  
\bibitem{Hoffman_Turner}
  M.~B.~Hoffman and M.~S.~Turner,
  Phys.\ Rev.\  D {\bf 64}, 023506 (2001).

\bibitem{Kinney}
  W.~H.~Kinney,
  Phys.\ Rev.\  D {\bf 66}, 083508 (2002).

\bibitem{Easther_Kinney}
  R.~Easther and W.~H.~Kinney,
  Phys.\ Rev.\  D {\bf 67}, 043511 (2003).
  
\bibitem{Liddle_flow}
  A.~R.~Liddle,
  Phys.\ Rev.\ D {\bf 68}, 103504 (2003).
  
\bibitem{Kinney:2006qm}
  W.~H.~Kinney, E.~W.~Kolb, A.~Melchiorri and A.~Riotto,
  Phys.\ Rev.\  D {\bf 74}, 023502 (2006).

\bibitem{Kinney:2002qn}
  W.~H.~Kinney,
  Phys.\ Rev.\  D {\bf 66}, 083508 (2002).
  
\bibitem{Kinney:2005in}
  W.~H.~Kinney and A.~Riotto,
  JCAP {\bf 0603}, 011 (2006).

\bibitem{Cortes:2007ak}
  M.~Cortes, A.~R.~Liddle and P.~Mukherjee,
  arXiv:astro-ph/0702170.

\bibitem{Peiris:2006sj}
  H.~Peiris and R.~Easther,
  JCAP {\bf 0610}, 017 (2006).

\bibitem{Shiu}
  G.~Shiu and B.~Underwood,
  Phys.\ Rev.\ Lett.\  {\bf 98}, 051301 (2007);
  S.~Kecskemeti, J.~Maiden, G.~Shiu and B.~Underwood,
  JHEP {\bf 0609}, 076 (2006).

\bibitem{Hinshaw:2006ia}
  G.~Hinshaw {\it et al.}  [WMAP Collaboration],
  arXiv:astro-ph/0603451.
 
\bibitem{Page:2006hz}
  L.~Page {\it et al.}  [WMAP Collaboration],
  arXiv:astro-ph/0603450.
  
\bibitem{Tegmark:2003uf}
  M.~Tegmark {\it et al.}  [SDSS Collaboration],
  Astrophys.\ J.\  {\bf 606}, 702 (2004).
    
\bibitem{Peiris:2006ug}
  H.~Peiris and R.~Easther,
  JCAP {\bf 0607}, 002 (2006).
  
\bibitem{Komatsu:2001rj}
  E.~Komatsu and D.~N.~Spergel,
  Phys.\ Rev.\  D {\bf 63}, 063002 (2001).

\bibitem{Verde}
  L.~Verde, H.~Peiris and R.~Jimenez,
  JCAP {\bf 0601}, 019 (2006).

\bibitem{Kesden}
  M.~Kesden, A.~Cooray and M.~Kamionkowski,
  Phys.\ Rev.\ Lett.\  {\bf 89}, 011304 (2002).

\bibitem{Knox}
  L.~Knox and Y.~S.~Song,
  Phys.\ Rev.\ Lett.\  {\bf 89}, 011303 (2002).

\bibitem{DBIv2Eva}
Eva Silverstein, private communication.

\bibitem{DBIv2SarahLouis}
Sarah Shandera and Louis Leblond, private communications.

\bibitem{Burgess}
  C.~P.~Burgess, J.~M.~Cline, K.~Dasgupta and H.~Firouzjahi,
  arXiv:hep-th/0610320.
  
\bibitem{Krause}
  A.~Krause and E.~Pajer,
  arXiv:0705.4682 [hep-th].

\end{thebibliography}
\end{document}